\documentclass[10pt,journal,compsoc]{IEEEtran}
\pdfoutput=1
\usepackage{amsmath,amssymb,amsfonts,amsthm}
\usepackage{xspace}
\usepackage{epstopdf}
\usepackage{pstricks}
\usepackage{times}
\usepackage{graphicx}
\usepackage{subfigure}
\usepackage{float}
\usepackage{enumitem}
\usepackage{caption}
\usepackage{url}
\usepackage{multirow}
\usepackage{multicol}

\usepackage{balance}
\usepackage[ruled,vlined,linesnumbered,noend]{algorithm2e}
\usepackage{caption}
\usepackage{balance}
\usepackage{circledsteps}
\usepackage{xcolor}
\usepackage{colortbl}
\usepackage{slashed}
\usepackage{subfigure}
\usepackage{cite}
\usepackage{hyperref}
\usepackage[numbers,sort&compress]{natbib}

\usepackage{atbegshi}
\AtBeginDocument{\AtBeginShipoutNext{\AtBeginShipoutDiscard}\addtocounter{page}{-1}}

\definecolor{Gray}{gray}{0.90}
\definecolor{Yellow}{rgb}{1,1,0.5}
\definecolor{Cyan}{rgb}{0.9,1,1}
\definecolor{Green}{rgb}{0.93,1,0.93}
\newcommand{\cg}{\cellcolor{Gray}}

\newcounter{example}[section]
\renewcommand{\theexample}{\nthesection.\arabic{example}}
\newenvironment{example}{
     \refstepcounter{example}
     {\vspace{1ex} \noindent\bf  Example  \theexample:}}{
     \eop\vspace{0.5ex}} 

\newcounter{definition}[section]
\renewcommand{\thedefinition}{\nthesection.\arabic{definition}}
\newenvironment{definition}{
     \refstepcounter{definition}
     {\vspace{1ex} \noindent\bf  Definition  \thedefinition:}}{
     \eop\vspace{0.5ex}} 

\newcounter{theorem}[section]
\renewcommand{\thetheorem}{\nthesection.\arabic{theorem}}
\newenvironment{theorem}{\begin{em}
        \refstepcounter{theorem}
        {\vspace{1ex} \noindent\bf  Theorem  \thetheorem:}}{
        \end{em}\eop\vspace{0.5ex}} 

\newcounter{lemma}[section]
\renewcommand{\thelemma}{\nthesection.\arabic{lemma}}
\newenvironment{lemma}{\begin{em}
        \refstepcounter{lemma}
        {\vspace{1ex}\noindent\bf Lemma \thelemma:}}{
        \end{em}\eop\vspace{0.5ex}} 

\newcounter{remark}[section]
\renewcommand{\theremark}{\nthesection.\arabic{remark}}

\newcommand{\myproof}{\noindent{\emph {Proof:} }}

\newcommand{\nthesection}{\arabic{section}}

\newcommand{\eop}{\hspace*{\fill}\mbox{$\Box$}}

\newcommand{\stitle}[1]{\vspace{1ex} \noindent{{\bf #1}}}

\newcommand{\sstitle}[1]{\vspace{1ex} \noindent{\textit{ #1}}}

\newcommand{\kw}[1]{{\ensuremath {\mathsf{#1}}}\xspace}

\newcommand{\kwnospace}[1]{{\ensuremath {\mathsf{#1}}}}

\newcommand{\ei}{\end{itemize}}
\newcommand{\ee}{\end{enumerate}}

\newcommand{\beqn}{\begin{eqnarray*}}
\newcommand{\eeqn}{\end{eqnarray*}}

\newcounter{ccc}

\newcommand{\eat}[1]{}

\newfloat{tcm}{thp}{loa}
\floatname{tcm}{Recursive \sql}

\newcommand{\sql}{{\sc sql}\xspace}

\long\def\comment#1{}

\definecolor{lgray}{gray}{0.85}
\definecolor{llgray}{gray}{0.9}

\newcommand{\reffig}[1]{Fig.~\ref{fig:#1}}
\newcommand{\refsec}[1]{Section~\ref{sec:#1}}
\newcommand{\reftable}[1]{Table~\ref{tab:#1}}
\newcommand{\refalg}[1]{Algorithm~\ref{alg:#1}}

\newcommand{\refdef}[1]{Definition~\ref{def:#1}}
\newcommand{\refthm}[1]{Theorem~\ref{thm:#1}}
\newcommand{\reflem}[1]{Lemma~\ref{lem:#1}}

\makeatletter

\newcommand{\Rmnum}[1]{\expandafter\@slowromancap\romannumeral #1@}
\makeatother

\newcommand{\nbr}{\kw{nbr}\xspace}

\newcommand{\mydeg}{\kw{deg}}

\newcommand{\scan}{\kw{SCAN}}

\newcommand{\gpuscan}{\kw{GPUSCAN}}
\newcommand{\gpuscanp}{\kw{GPUSCAN^{++}}}
\newcommand{\gpuscanpo}{\kw{GPUSCAN^{++}_O}}
\newcommand{\gpuscanpuvm}{\kw{GPUSCAN^{++}_{UVM}}}

\begin{document}

\title{{$\textrm{GPUSCAN}^{++}$}: Efficient Structural Graph Clustering on GPUs
}


\author{
 Long Yuan, Zeyu Zhou, Xuemin Lin, Zi Chen, Xiang Zhao, Fan Zhang
}

\thanks{* Zi Chen is the corresponding author of this paper.}

\IEEEcompsocitemizethanks{

\IEEEcompsocthanksitem L. Yuan and Z. Zhou are with the School of Computer Science and Engineering, Nanjing University of Science and Technology, Nanjing, China.\protect\\
E-mail: \{zeyuzhou,longyuan\}@njust.edu.cn.

\IEEEcompsocthanksitem X. Lin is with the School of Computer Science and Engineering, the University of New South Wales, Sydney, Australia.\protect\\
E-mail: lxue@cse.unsw.edu.au.

\IEEEcompsocthanksitem Z. Chen is with the Software Engineering Institute, East China Normal University, Shanghai, China.\protect\\
E-mail: zchen@sei.ecnu.edu.cn.

\IEEEcompsocthanksitem X. Zhao is with the College of Systems Engineering, National University of Defense Technology, Changsha, China.\protect\\
E-mail: xiangzhao@nudt.edu.cn.

\IEEEcompsocthanksitem F. Zhang is with Cyberspace Institute of Advanced Technology, Guangzhou University, Guangzhou, China.\protect\\
E-mail: fanzhang.cs@gmail.com.


}
\thanks{Manuscript received xxx; revised xxx.}

%

\IEEEtitleabstractindextext{
\begin{abstract}
 
 Structural clustering is one of the most popular graph clustering methods, which has achieved great performance improvement by utilizing GPUs.  Even though, the state-of-the-art GPU-based structural clustering algorithm, \gpuscan, still suffers from efficiency issues since lots of extra costs are introduced for parallelization. Moreover,  \gpuscan assumes that the graph is resident in the GPU memory. However, the GPU memory capacity is limited currently while  many real-world graphs are big and cannot fit in the GPU memory, which makes \gpuscan unable to handle large graphs. Motivated by this, we present a new GPU-based structural clustering algorithm, \gpuscanp, in this paper. To address the efficiency issue, we propose a new progressive clustering method tailored for GPUs that not only avoid high parallelization costs but also fully exploits the computing resources of GPUs.  To address the GPU memory limitation issue, we propose a partition-based algorithm for structural clustering that can  process large graphs with limited GPU memory. We conduct experiments on real graphs, and the experimental results demonstrate that our algorithm can achieve up to 168 times speedup compared with the state-of-the-art GPU-based algorithm when the graph can be resident in the GPU memory. Moreover, our algorithm is scalable to handle large graphs. As an example, our algorithm can finish the structural clustering on a graph with 1.8 billion edges using less than 2 GB GPU memory.
\end{abstract}

}

\maketitle

\IEEEdisplaynontitleabstractindextext
\IEEEpeerreviewmaketitle

\section{Introduction}


Graph clustering is one of the most fundamental problems in analyzing graph data. It  has been extensively studied \cite{guimera2005functional,palla2005uncovering,schaeffer2007graph,xu2007scan,yin2017local} and frequently utilized in many applications, such as natural language processing \cite{biemann2006chinese}, recommendation systems \cite{bellogin2012using}, social and biological network analysis \cite{girvan2002community,martha2011constructing}, and load balancing in distributed systems\cite{fan2020application}.

\begin{figure}[!tbp]
	\centerline{\includegraphics[width=0.42\textwidth]{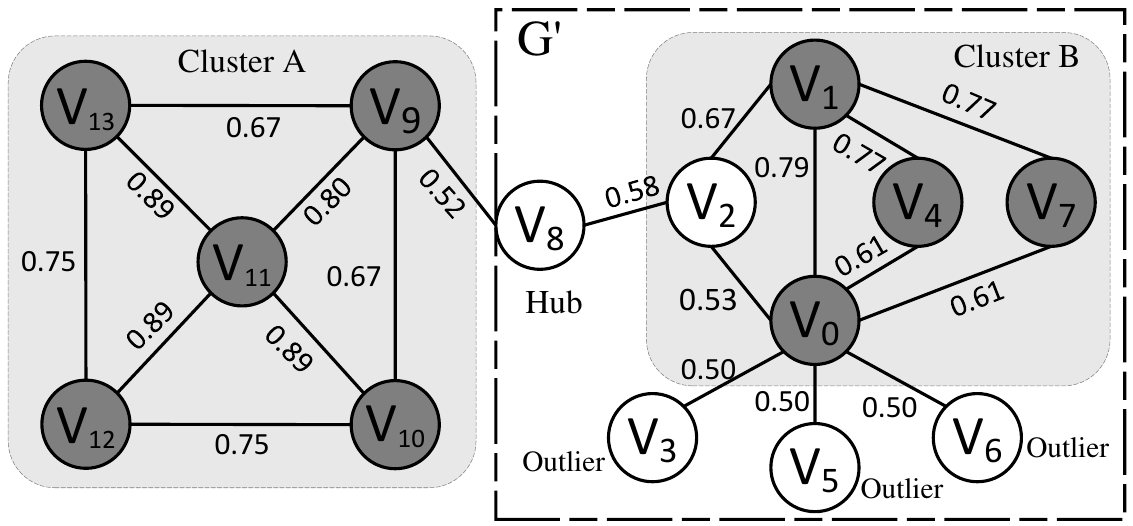}}
	\caption{Clustering result of \scan ($\epsilon = 0.6, \mu=3$) on $ G $}
	\vspace{-0.2cm}
	\label{fig:graph}
	\vspace{-0.4cm}
\end{figure}

One well-known method of graph clustering is structural clustering, which is introduced via the Structural Clustering Algorithm for Networks (\kw{SCAN}) \cite{xu2007scan}. Structural clustering defines the structural similarity between two vertices, and two vertices with structural similarity not less than a given parameter $\epsilon$ are considered similar. A core is a vertex that has at least $\mu$ similar neighbors, where $\mu$ is also a given parameter. A cluster contains the cores and vertices that are structurally similar to the cores.  If a vertex does not belong to any cluster but its neighbors belong to more than one cluster, then it is a hub; Otherwise, it is an outlier. \reffig{graph} shows the clustering result of structural clustering with $\epsilon = 0.6$ and $\mu = 3$ on graph $G$ in which two clusters are colored gray, cores are colored black, and hubs and outliers are labeled. Compared with most graph clustering approaches aiming to partition the vertices into disjoint sets, structural clustering is also able to find \emph{hub} vertices that connected different clusters and \emph{outlier} vertices that lack strong ties to any clusters, which is important for mining various complex  networks \cite{xu2007scan,DBLP:conf/icdm/KangF11,DBLP:conf/sigmod/TsengDS21}.  Structural clustering has been successfully used to cluster biological data \cite{ding2012atbionet,martha2011constructing} and social media data \cite{martha2013study,li2015social,schinas2015visual,schinas2015multimodal}. 


%

\stitle{Motivation.} In recent years, the rapid evolution of Graphics Processing Unit (GPU) has attracted extensive attention in both industry and academia, and it has been used  successfully to speed up various graph algorithms \cite{lin2016network,hu2021accelerating,ye2021gpu,sha2021self,shi2018graph}.   The massively parallel hardware resources in GPUs also offer promising opportunities for accelerating structural clustering. In the literature, a GPU-based algorithm named \gpuscan is proposed \cite{stovall2014gpuscan}. By redesigning the computation steps of structural clustering, \gpuscan significantly improves the clustering performance compared with the state-of-the-art serial algorithm. However, it has the following drawbacks:
 
\begin{itemize}[leftmargin=*]
\item \emph{Prohibitive parallelization overhead.} \gpuscan aims to fully utilize the massively parallel hardware resources in GPU. However, lots of extra  computation cost is introduced for this goal. Given a graph $G=(V, E)$, the state-of-the art serial  algorithm \cite{Chang2017pSCAN} can finish the clustering with $O(m \cdot \mydeg_{\kw{max}})$ work while \gpuscan needs $ O(\Sigma_{(u, v) \in E(G)} (\mydeg(u)+\mydeg(v))+ c \cdot m \cdot \log m)  $  work  \footnote{We use the work-span framework to analyze the parallel algorithm \cite{cormen2022introduction}. The work of an algorithm is the number of operations it performs. The span of an algorithm is the length of its longest sequential dependence.} for it, where $m$ denotes the number of edges in the graph, $\mydeg_{\kw{max}}$ denotes the maximum degree of the vertices in $G$, and $c$ is a not small integer as verified in our experiments determined by the given graph $G$ and parameter $\epsilon$ and $\mu$.  The prohibitive parallelization overhead makes \gpuscan  inefficient for structural clustering. 

\item \emph{GPU memory capacity limitation.} Compared to common server machines equipped with hundreds of gigabytes or even terabytes of main memory, the GPU memory capacity is very limited currently. \gpuscan assumes that the graph is resident in the GPU memory. Nevertheless, many real-world graphs are big and may not fit entirely in the GPU memory. The memory capacity limitation significantly restricts the scale of graphs that \gpuscan can handle.
\end{itemize}

Motivated by this, in this paper, we aim to devise a new efficient GPU-based structural clustering algorithm that can handle large graphs.


\stitle{Our solution.} We address the drawbacks of \gpuscan and propose new GPU-based algorithms for structural clustering in this paper. Specifically, for the prohibitive parallelization overhead, we devise a CSR-enhanced graph storage structure and  adopt a progressive clustering paradigm with which unnecessary similarity computation can be pruned. Following this paradigm, we develop a new GPU-based \scan algorithm with careful considerations of total work and parallelism, which are crucial to the performance of GPU programs. Our new algorithm significantly reduces the parallelization overhead. We theoretically prove that the work of our new algorithm is reduced to $O(\Sigma_{(u, v) \in E(G)}\mydeg(u) \cdot \log \mydeg(v))$, and experimental results show that our algorithms achieve 85 times speedup on average compared with \kwnospace{GPUSCAN}.

For the GPU memory capacity limitation, a direct solution is to use the Unified Virtual Memory (UVM) provided by recent GPUs. UVM utilizes the large host main memory of common server machines and allows GPU memory to be oversubscribed as long as there is sufficient host main memory to back the allocated memory. The driver and the runtime system handle data movement between the host main memory and GPU memory without the programmer's involvement, which allows running GPU applications that may otherwise be infeasible due to the large size of datasets \cite{gera2020traversing}. However, the performance of this approach is not competitive due to poor data locality caused by irregular memory accesses during the structural clustering and relatively slow rate I/O bus connecting host memory and GPU memory. Therefore, we propose a new partition-based structural clustering algorithm. Specifically, we partition the graph into several specified subgraphs. When conducting the clustering, we only need to load the specified subgraphs into the GPU memory and we can guarantee that the clustering results are the same as loading the whole graph in the GPU memory, which not only addresses the GPU memory limitation problem in \kwnospace{GPUSCAN} but also avoids the data movement overheads involved in the UVM-based approach.


\stitle{Contributions.} We make the following contributions:

\sstitle{(A)} We theoretically analyze the performance of the state-of-the-art GPU-based structural clustering  algorithm \gpuscan and reveal the key reasons leading to its inefficiency. Following the analysis,  we propose a new progressive clustering algorithm tailored for GPUs that not only avoids the prohibitive parallelization overhead but also fully exploits the computing power of GPUs.

\sstitle{(B)} To overcome the GPU memory capacity limitation, we devise a new out-of-core GPU-based SCAN algorithm. By partitioning the original graph into a series of smaller subgraphs, our algorithm can significantly reduce the GPU memory requirement when conducting the clustering and scale to large data graphs beyond the GPU memory.

\sstitle{(C)} We conduct extensive performance studies using ten real graphs. The experimental results show that our proposed algorithm achieves 168/85 times speedup at most/on average,  compared with the state-of-the-art GPU-based SCAN algorithm. Moreover, our algorithm can finish the structural clustering on a graph with 1.8 billion edges using less than 2 GB of GPU memory.

\section{Preliminaries}
\label{sec:pre}

\subsection{Problem Definition}

Let $G=(V,E)$ be an undirected and unweighted graph, where $V(G)$ is the set of vertices and $E(G)$ is the set of edges, respectively. We denote the number of vertices as $n$ and the number of edges as $m$, i.e., $n = |V(G)|$, $m = |E(G)|$. For a vertex $v \in V(G)$, we use $\nbr(v, G)$ to denote the set of neighbors of $v$. The degree  of a vertex $v \in V(G)$, denoted by $\mydeg(v, G)$ is the number of neighbors  of $v$, i.e., $\mydeg(v, G) = |\nbr(v, G)|$. For simplicity, we omit $G$ in the notations when the context is self-evident.

\begin{definition}
\label{def:structuraln} 
\textbf{(Structural Neighborhood)} Given a vertex $v$ in $G$, the structural neighborhood of $v$, denoted by $N[v]$, is defined as $N[v]=\{u \in V(G) |(u, v) \in E(G) \} \cup \{v\}$.
\end{definition}

\begin{definition}
\label{def:ss}
\textbf{(Structural Similarity)} Given two vertices $u$ and $v$ in $G$, the structural similarity between $u$ and $v$, denoted by $\sigma(u, v)$, is defined as:
	\begin{equation}
		\sigma(u,\;v)=\frac{\vert N\lbrack u\rbrack\cap N\lbrack v\rbrack\vert}{\sqrt{\vert N\lbrack u\rbrack\vert\vert N\lbrack v\rbrack\vert}}\label{eq}
	\end{equation}
\end{definition}



\begin{definition}
\label{def:epsN}
\textbf{($\epsilon$-Neighborhood)}
Given a similarity threshold $0<\epsilon\leq 1$, the $\epsilon$-neighborhood of $u$, denoted by $ N_\epsilon\lbrack u\rbrack $, is defined as the subset of $N[u]$ in which every vertex $v$ satisfies $\sigma(u, v) \geq \epsilon$, i.e., $ N_\epsilon\lbrack u\rbrack=\{v\vert v\in N\lbrack u\rbrack\wedge\sigma(u,v) \geq \epsilon\}$. 
\end{definition}

Note that the $\epsilon$-neighborhood of a given vertex $u$ contains $u$ itself as $\sigma(u, u) = 1$. When the number of $\epsilon$-neighbors of a vertex is large enough, it becomes a core vertex: 

\begin{definition}
\textbf{(Core)} Given a structural similarity threshold $0 < \epsilon \leq 1$ and an integer $\mu \geq 2$, a vertex $u$ is a core vertex if $ \vert N_\varepsilon[u]\vert\geq\mu $.
\end{definition}

Given a core vertex $u$, the structurally reachable vertex of $u$ is defined as:

\begin{definition}
\textbf{(Structural Reachability)} Given two parameters $ 0< \epsilon \leq1 $ and $\mu \geq 2$, for two vertices $u$ and $v$,  $v$ is structurally reachable from vertex $ u $ if there is a sequence of vertices $ v_1,v_2,...,v_l \in V $ such that: (i) $ v_1=u, v_l=v $; (ii) $v_1, v_2,...,v_{l-1} $ are core vertices; and (iii) $ v_{i+1}\in N_\epsilon\lbrack v_i\rbrack $ for each $ 1\leq i\leq l-1 $.
\end{definition}


Intuitively, a cluster is a set of vertices that can be structurally reachable from any core vertex. Formally,

\begin{definition}\label{def:cluster} \textbf{(Cluster)} A cluster $C$ is a non-empty subset of $V$ such that:
	\begin{itemize}[leftmargin=*]
		\item (Connectivity) For any two vertices $v_1, v_2 \in C$, there exists a vertex $u \in C$ such that both $v_1$ and $v_2$ are structurally reachable from $u$.
		\item (Maximality) If a core vertex $u \in C$, then all vertices which are structure-reachable from $u$ are also in $C$.
	\end{itemize}
\end{definition}

\begin{definition}
	\textbf{(Hub and Outlier)} Given the set of clusters in graph $G$, a vertex $u$  not in any cluster is a hub vertex if its neighbors belong to two or more clusters, and it is an outlier vertex otherwise. 
\end{definition}

Following the above definitions, it is clear that two vertices that are not adjacent in the graph have no effects on the clustering results even if they are similar. Therefore, we only need to focus on the vertex pair incident to an edge. In the following,  \emph{for an edge $(u, v)$, we use the edge similarity of $(u, v)$ and the similarity of $u$ and $v$ \underline{interchangeably}  for brevity}. We summarize the notations used in the paper in \reftable{notations}.

\begin{table}[t]
\renewcommand{\arraystretch}{1.3}
\centering
{\scriptsize
\begin{tabular}{r l}
 \hline
  Symbol & Description  \\
  \cline{1-2}
  $G = (V, E) $&  {\scriptsize Graph with vertices $V$ and edges $E$ }     \\
   $V(G)/E(G)$&   {\scriptsize All vertices/edge in $G$}   \\
   $\nbr(u)/\mydeg(u)$ &  {\scriptsize Neighbors/degree of a vertex u}   \\
   $n/m$ & {\scriptsize number of vertices/edges in $G$}\\	
   $\mydeg_{\kw{max}}$ & {\scriptsize maximum degree in $G$} \\
   \vspace{0.1cm}
    $\normalsize\textcircled{\scriptsize{\bf ?}}$ & {\scriptsize the role of a vertex or the similarity of an edge is unknown}\\
    $\normalsize\textcircled{\scriptsize{\bf C}}$\normalsize{/}$ \normalsize\textcircled{\scriptsize{\bf !C}}$ & {\scriptsize  the vertex is a core/non-core vertex }\\
    $\normalsize\textcircled{\scriptsize{\bf S}}$\normalsize{/}$\normalsize\textcircled{\scriptsize{\bf !S}}$ & {\scriptsize the edge is similar/dis-similar}\\
    $\normalsize\textcircled{\scriptsize{\bf H}}$\normalsize{/}$\normalsize\textcircled{\scriptsize{\bf O}}$ & {\scriptsize  the vertex is a hub/outlier} \\
   
     \hline
\end{tabular}
\caption{\small{Notations}}
\label{tab:notations}
\vspace*{-0.4cm}
}
\end{table}

%

\begin{example}
Considering $G$ shown in \reffig{graph} where $\epsilon = 0.6$ and $\mu = 3$, the structural similarity of every pair of adjacent vertices is shown near the edge. $v_0$, $v_1$, $v_4$, $v_7$, $v_9$, $v_{10}$, $v_{11}$, $v_{12}$, and $ v_{13} $ are core vertices as they all have at least $\mu-1=2$ similar neighbors including themselves. These core vertices form two clusters, which are marked in grey. $v_8$ is a hub as its neighbors $v_2$ and $v_9$ are in different clusters, while $v_3,v_5$ and $ v_6 $ are outliers. Due to the limited space, in the following examples, we only show procedures on the subgraph $G'$ of $G$.
\end{example}

\stitle{Problem Statement.} Given a graph $G=(V, E)$ and two parameters $ 0< \epsilon \leq 1$ and $\mu \geq 2$, structural graph clustering aims to efficiently compute all the clusters $\mathbb{C}$ in $G$ and  identify the corresponding role of each vertex. In this paper, we aim to accelerate the clustering performance of \scan by GPUs.

\subsection{GPU Architecture and CUDA Programming Platform}


\stitle{GPU Architecture.}  A GPU consists of multiple streaming multiprocessors (SM). Each SM has a large number of GPU cores. The cores in each SM work in single-instruction multiple-data (SIMD) \footnote{Single-instruction multi-thread (SIMT) model in NIVIDA's term} fashion, i.e., they execute the same instructions at the same time on different input data. The smallest execution units of a GPU are \emph{warps}. In NVIDIA's GPU architectures, a warp typically contains 32 threads. Threads in the same warp that finish their tasks earlier have to wait until other threads finish their computations, before swapping in the next warp of threads. A GPU is equipped with several types of memory. All SMs in the GPU share a larger but slower \emph{global memory} of the GPU. Each SM has small but fast-access local programmable memory (called \emph{shared memory}) that is shared by all of its cores. Moreover, data can be exchanged between a GPU's global memory and the host's main memory via a relatively slow rate I/O bus.

\stitle{CUDA.} CUDA (Compute Unified Device Architecture) is the most popular GPU programming language. A CUDA problem consists of codes running on the CPU (\emph{host code}) and those on the GPU (\emph{kernel}). Kernels are invoked by the host code. Kernels start by transferring input data from the main memory to the GPU's global memory and then processing the data on the GPU. When finishing the process, results are copied from the GPU's global memory back to the main memory. The GPU executes each kernel with a user-specified number of threads. At runtime, the threads are divided into several \emph{blocks} for execution on the cores of an SM, each block contains several warps, and each warp of threads executes concurrently.

For the thread safety of GPU programming, CUDA provides programmers with atomic operations, including \kw{atomicAdd} and \kw{atomicCAS}. \kw{atomicAdd(\emph{address},\emph{val})}reads the 16-bit, 32-bit, or 64-bit word \emph{old} located at the address \emph{address} in global or shared memory, computes \kw{(\emph{old}+\emph{val})}, and stores the result back to memory at the same address. These three operations are performed in one atomic transaction and returns \emph{old} value.  \kw{atomicCAS(\emph{address},\emph{old},\emph{new})} (Compare And Swap) reads the 16-bit, 32-bit, or 64-bit word \emph{old} located at the address \emph{address} in global or shared memory, computes  \kw{(\emph{old}==\emph{old}?\emph{new}:\emph{old})}, and stores the result back to memory at the same address. These three operations are performed in one atomic transaction and return \emph{old}.


\section{Related work}
\label{sec:existing}

\subsection{CPU-based \scan}

Structural graph clustering (\scan) is first proposed in \cite{xu2007scan}. After that, a considerable number of follow-up works are conducted in the literature. For the online algorithms regarding a fixed parameter $\epsilon$ and $\mu$, \kwnospace{SCAN}++ \cite{shiokawa2015scan++}, pSCAN \cite{Chang2017pSCAN} and  \kw{anySCAN} \cite{DBLP:conf/icdm/ZhaoCX17} speed up \scan by pruning unnecessary structural similarity computation. \kw{SparkSCAN} \cite{zhou2015sparkscan} conducts the structural clustering on the Spark system. \kw{pmSCAN} \cite{DBLP:conf/cikm/SeoK17} considers the case that the input graph is stored on the disk. \kw{ppSCAN} \cite{DBLP:conf/icpp/CheSL18} parallelizes pruning-based algorithms and uses vectorized instructions to further improve the performance. 

For the index-based algorithms aiming at returning the clustering result fast for frequently issued queries, GS*-Index \cite{wen2017efficient} designs a space-bounded index and proposes an efficient algorithm to answer a clustering query.   \cite{DBLP:conf/sigmod/TsengDS21} parallelizes  GS*-Index and uses locality-sensitive hashing to design a novel approximate index construction algorithm. \cite{DBLP:conf/sigmod/RuanGWW21} studies the maintenance of computed structural similarity between vertices when the graph is updated and proposes an approximate algorithm that achieves $O(\log^2|V|+\log|V|\cdot\log\frac{M}{\delta^*})$ amortized cost for each update for a specified failure probability $\delta^*$ and every sequence of $M$ updates. \cite{lingkaimeng2022} studies structural clustering on directed graphs and proposes an index-based approach to computing the corresponding clustering results. However, all these algorithms are CPU-based and cannot be translated into efficient solutions on GPUs due to the inherently unique characteristics of a GPU different from a CPU.

Besides \scan,  other graph clustering methods are also proposed in the literature. However, all these methods are not based on structural clustering, and thus do not share the effectiveness of \scan in detecting clusters, hubs, and outliers. \cite{schaeffer2007graph,aggarwal2010survey} provides a comprehensive survey on this topic.

\subsection{GPU-based \scan}
\label{sec:gpuscan}

For the GPU-based approach, \kwnospace{GPUSCAN} \cite{DBLP:journals/tpds/StovallKA15} is the state-of-the-art GPU-based \scan solution, which conducts the  clustering in three separate phases: (1)  $\epsilon$-neighborhood identification. In Phase 1, \scan computes the  similarity of each edge and determines the core vertices in the graph. (2) cluster detection.  After identifying the core vertices,  \kwnospace{GPUSCAN}  detects the  clusters based on \refdef{cluster} in \refsec{pre}.  (3) hub and outlier classification. With the clustering results, the hub and outlier vertices are classified  in Phase 3. The following example illustrates the three phases of \gpuscan.

\begin{figure}[t]
	\centering
	\subfigure[Phase 1: $\epsilon$-neighborhood identification]{
		\begin{minipage}[b]{0.48\textwidth}
			\includegraphics[width=1\textwidth]{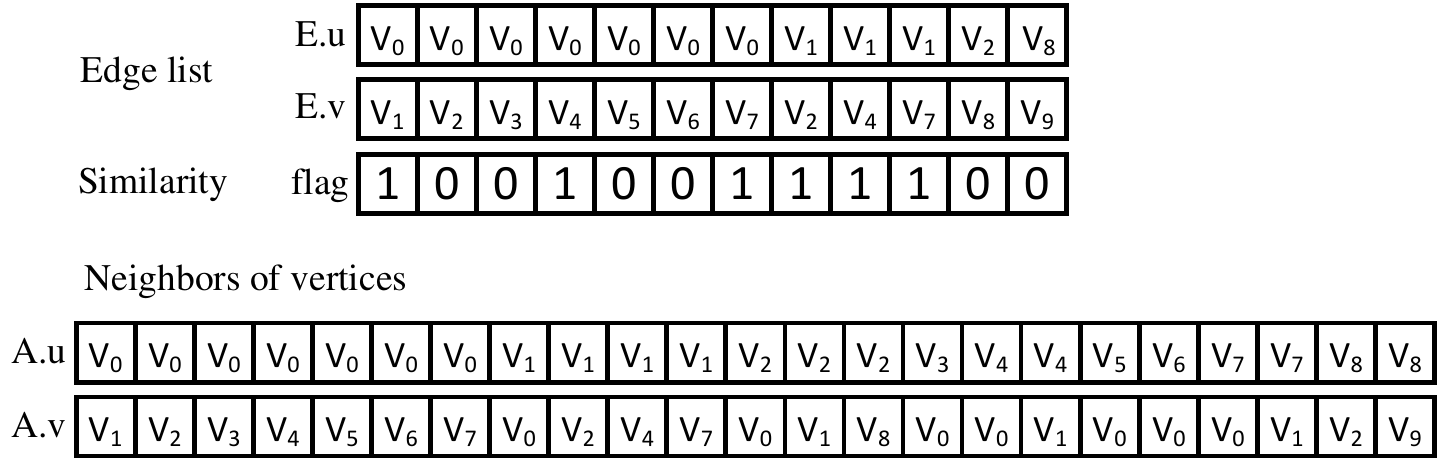}
			\vspace{-0.4cm}
		\end{minipage}
	}
	\vspace{-0.2cm}
	\subfigure[Phase 1: Compute common neighbors of $ v_0 $ and $ v_1 $]{
		\begin{minipage}[b]{0.48\textwidth}
			\includegraphics[width=1\textwidth]{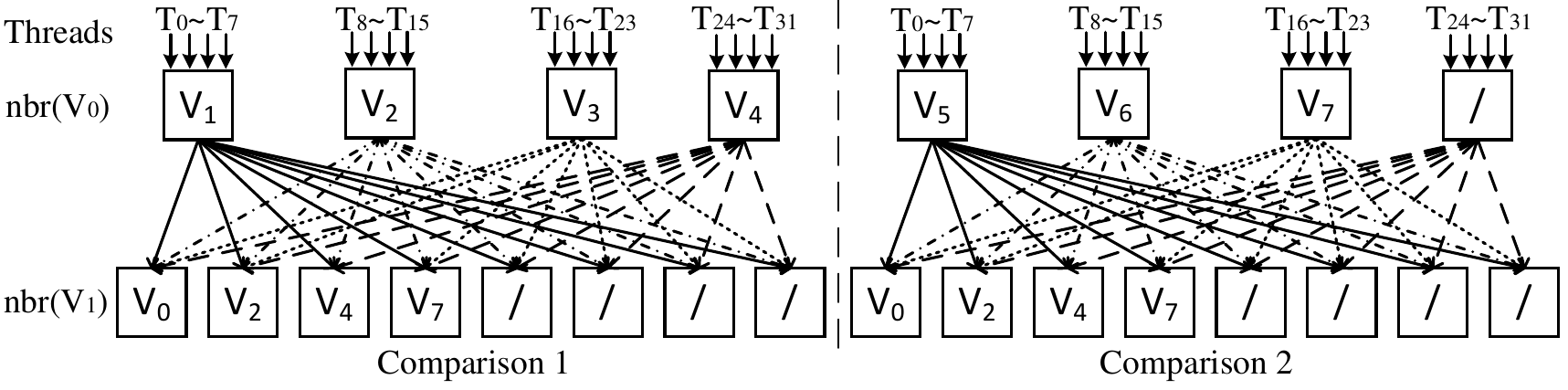}
			\vspace{-0.4cm}
		\end{minipage}
	}
	\vspace{-0.2cm}
	\subfigure[Phase 2: cluster detection]{
		\begin{minipage}[b]{0.48\textwidth}
			\includegraphics[width=1\textwidth]{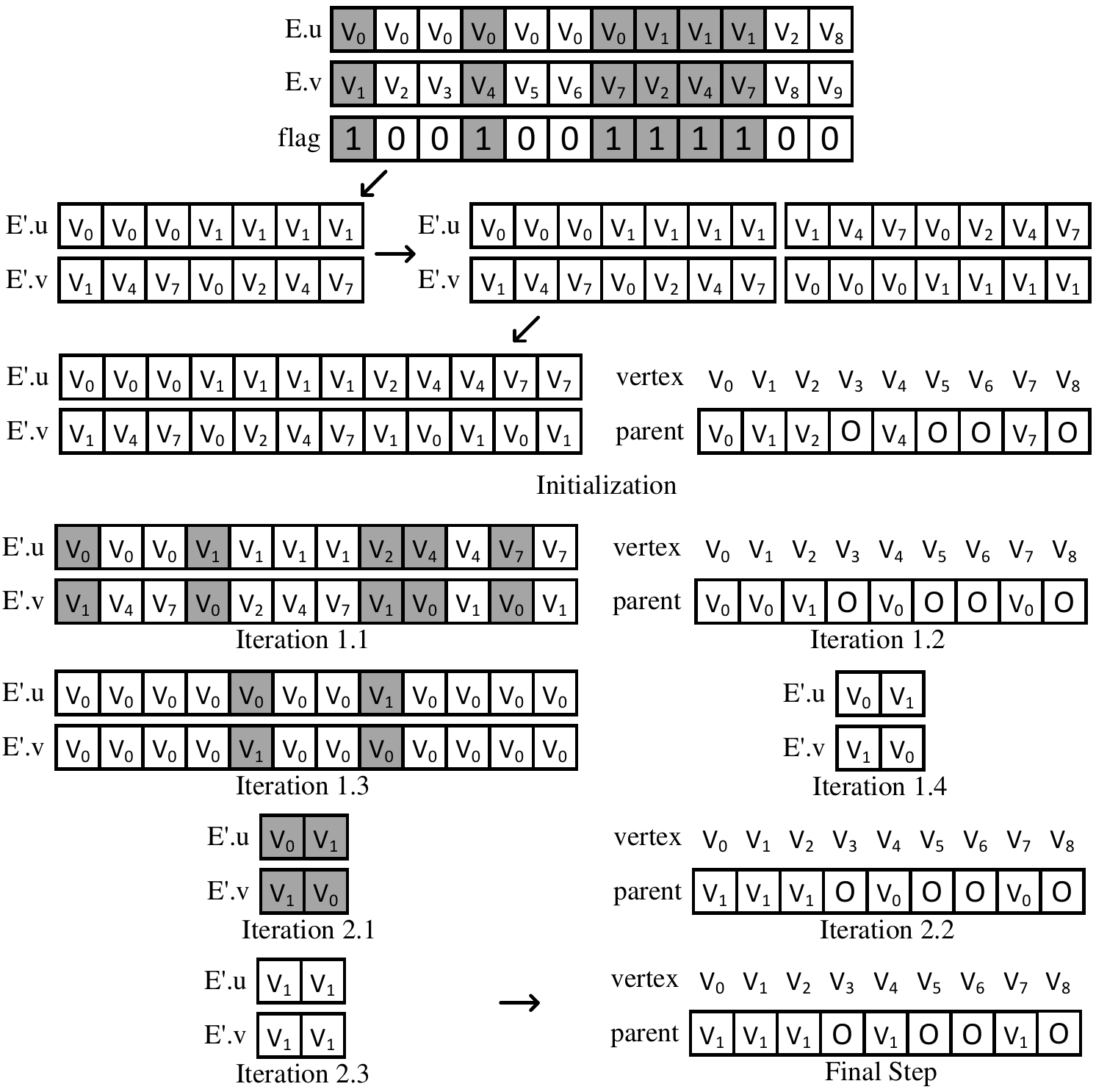}
			\vspace{-0.4cm}
		\end{minipage}
	}
	\vspace{-0.2cm}
	\subfigure[Phase 3: Hubs and outliers classification]{
		\begin{minipage}[b]{0.48\textwidth}
			\includegraphics[width=1\textwidth]{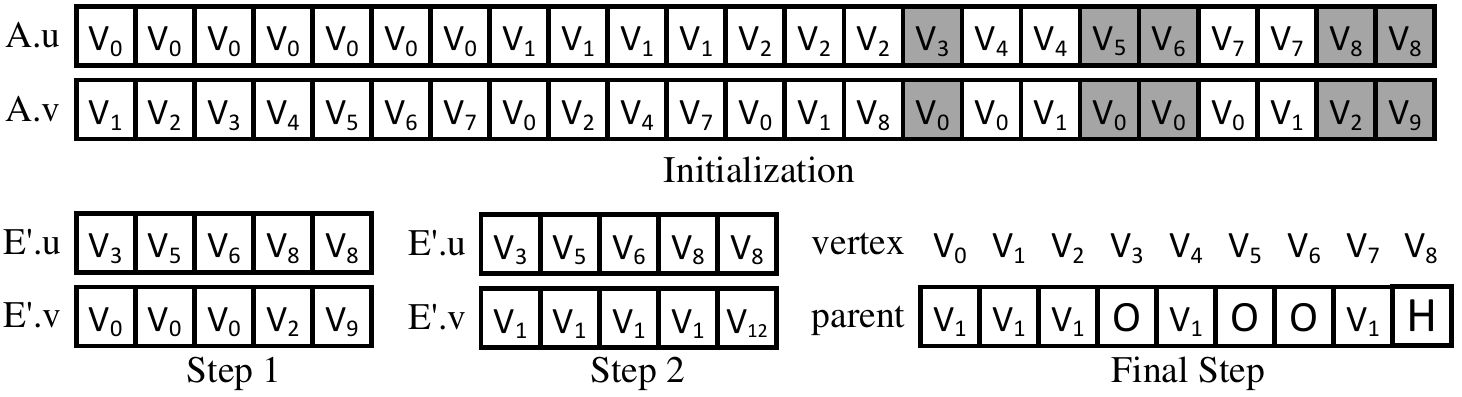}
			\vspace{-0.4cm}
		\end{minipage}
	}
	\caption{\gpuscan for vertices $G'$ ($\epsilon = 0.6, \mu=3$) } \label{fig:existgpuscan}
	\vspace{-0.4cm}
\end{figure}

\begin{example}
Consider graph $G$ shown in \reffig{graph}, \reffig{existgpuscan} shows the  clustering procedures of \kwnospace{GPUSCAN} with parameters $ \epsilon=0.6, \mu=3$. Due to the limited space, we only show the clustering procedures related to the subgraph $G'$ of $G$ in \reffig{graph} in the following examples and the remaining part can be processed similarly.

\reffig{existgpuscan} (a)-(b) show the $\epsilon$-neighborhood identification phase. \kwnospace{GPUSCAN}  uses  $E.u$ and $E.v$ arrays to represent the edges of the graph, a $\kw{flag}$ array in which each element stores whether an edge is similar or not, and $ A.u $ and $ A.v $ arrays to store the neighbors of each vertex.    \kwnospace{GPUSCAN}  uses a warp to computing the similarity of each edge $ (u,v) $. According to \refdef{ss}, the key to compute the similarity of $(u, v)$ is to find the common neighbors in $N[u]$ and $N[v]$.  To achieve this goal, for each $w \in \nbr(u)$, \kwnospace{GPUSCAN} uses 8 threads in a warp to check whether $w$ is also in 8 continuous elements in $\nbr(v)$. As a warp usually contains 32 threads, four elements in $\nbr(u)$ can be compared concurrently. After the comparisons of these four elements are complete, if the largest vertex id of these four elements in $\nbr(u)$ is bigger than the largest one in the 8 continuous elements  $ \nbr(v) $, then, the next continuous 8 elements in $ \nbr(v) $ are used to repeat the above comparisons. If the largest vertex id of these four elements in $\nbr(u)$ is smaller than the largest one in the 8 continuous elements  $ \nbr(v) $, then, the next 4 elements in $ \nbr(u)$ are used to repeat the above comparisons. Otherwise, the next 4 elements in $ \nbr(u) $ and the next 8 elements in $\nbr(v)$ are used to repeat the above comparisons. The common neighbors are found when all the elements in $\nbr(u)$ and $\nbr(v)$ have been explored.  Take edge $(v_0,v_1)$ as an example. \reffig{existgpuscan} (b) shows the procedure of \kwnospace{GPUSCAN}  to compute their common neighbors. As $\nbr(v_0) = \{v_1, v_2, v_3, v_4, v_5, v_6, v_7\}$, the first 4 elements $ \{v_1, v_2, v_3, v_4\} $ in $ \nbr(v_0) $ are compared first. For the 32 threads in a warp, threads $T_0-T_7$ handle the comparison for $ v_1 $. The remaining threads are assigned to handle the comparison for vertices $v_2$, $v_3$, and $v_4$ similarly. As $\nbr(v_1) = \{v_0, v_2, v_4, v_7\}$, threads $T_0-T_7$ finds $v_1 \notin \nbr(v_1)$. Similarly, threads $T_8-T_{15}$, $T_{16}-T_{23}$, $T_{24}-T_{31}$ find $v_2, v_4 \in \nbr(v_1)$. After that, they find the largest vertex id $ v_4 $ of elements in $ \nbr(v_0) $ that were compared is smaller than the largest one $ v_7 $ in $ \nbr(v_1) $. Then, the remaining 3 elements $ \{v_5, v_6, v_7\} $ in $ \nbr(v_0) $ are used to repeat the comparisons similarly. In this comparison, threads $T_{16}-T_{23}$ find that $ v_7 $ is the common element of $ \nbr(v_0) $ and $ \nbr(v_1) $. Therefore, $\sigma(v_0, v_1) = (2+3)/\sqrt{5*8} = 0.79 > 0.6$, which means $v_1$ and $v_0$ are similar, and the corresponding element in the \kw{flag} array is set as 1.
 
 


\reffig{existgpuscan} (c) shows the graph clustering phase of \kwnospace{GPUSCAN}. Following \refdef{cluster}, only similar edges could be in the final clusters. Therefore, \kwnospace{GPUSCAN} builds two new arrays $ E'.u $ and $ E'.v $ by retrieving the similar edges in $ E.u $ and $ E.v $. Consider the initialization step in \reffig{existgpuscan} (c), as $v_0$ and $v_1$ are similar while $v_0$ and $v_2$ are not similar, only $v_0$ and $v_1$ are kept in $E'.u$ and $E'.v$. Then Based on $E'.u$ and $E'.v$, $ N_\epsilon\lbrack u\rbrack $ can be obtained easily. For example, for $v_0$, $ N_\epsilon[v_0]  = \{v_0, v_1, v_4, v_7\}$, $| N_\epsilon\lbrack v_0\rbrack| = 4 > 3$, thus, $v_0$ is a core vertex. Similarly, $v_4$ and $v_7$ are also core vertices. Then, it removes the edges in $E'.u$ and $E'.v$ not incident to core vertices.  Moreover, \kw{GPUSCAN} represents all clusters by a spanning forest through  a \kw{parent} array. In the \kw{parent} array, each element represents the parent of the corresponding vertex in the forest,  and the corresponding element for the vertices in $E'.u$ and $E'.v$ is initialized as the vertex itself, which is also shown in Initialization of \reffig{existgpuscan} (c).

After that,  \kwnospace{GPUSCAN}  detects the clusters in the graph iteratively. In each odd iteration, for each vertex $u$ in $E'.u$, it finds the smallest neighbor $v$ of $u$ in  $E'.v$ and takes the smaller one between $u$ and $v$ as the parent of $u$ in the \kw{parent} array. Then, it replaces each vertex in $E'.u$ and $E'.v$ by its parent and removes the edges with the same incident vertices. Consider Iteration 1.1-1.4  in \reffig{existgpuscan} (c), for $v_1$, its smallest neighbor in $E'.v$ is $v_0$, which is smaller than itself, thus, the parent of $v_1$ is replaced by $v_0$ in the \kw{parent} array. The other vertices are processed similarly (Iteration 1.1 and 1.2). Then, $v_1$ is replaced by $v_0$ in $E'.u$ and $E'.v$ in Iteration 1.3, and only edges $(v_0, v_1)$ and $(v_1, v_0)$ are left after removing the edges with the same incident vertices in Iteration 1.4. In each even iteration, \kwnospace{GPUSCAN} conducts the clustering similarly to the odd iteration except the largest neighbor is selected as its parent. Iteration 2.1-2.4 in \reffig{existgpuscan} (c) show the even case. Note that to obtain the $E'.u$ and $E'.v$ at the end of each iteration, \kwnospace{GPUSCAN} uses \kw{sort()} and \kw{partition()} functions in the \kw{Thrust} library provided by Nvidia \cite{thrust}. When $E'.u$ and $E'.v$ become empty, \kwnospace{GPUSCAN} explores each vertex and sets its corresponding element in the \kw{parent} array as the root vertex in the forest. For example, in Iteration 2.2, the parent of $v_4$ is $v_0$, the parent of $v_0$ is $v_1$, and $v_1$ is the parent of itself, thus, the parent of $v_4$ is set as $v_1$. After Phase 2 finishes, the vertices with the same parent are in the same cluster. In  \reffig{existgpuscan} (c), as $ v_0 $, $ v_1 $, $ v_2 $, $ v_4 $, and $ v_7 $ have the same parent, they are in the same cluster.



In Phase 3, \kwnospace{GPUSCAN}  classifies the hub and outlier vertices. It first retrieves edge $(u, v)$ from $A.u$ and $A.v$ such that $u \in A.u$ is not in a cluster and $v \in A.v$ is in a cluster and keeps them in new arrays $E'.u$ and $E'.v$. Then, it replaces the vertices in $E'.v$ by their parents. Then, for a vertex $u \in E'.u$,  it has two more different neighbors, then it is a hub vertex. Otherwise, it is an outlier vertex. Consider the example shown in \reffig{existgpuscan} (d), $v_3$, $v_5$, $v_6$, and $v_8$ are not in a cluster, thus, the edges that they form with their neighbors within the cluster are retrieved and stored in $E'.u$ and $E'.v$ (Step 1). Then, $v_0$, $v_2$, and $v_9$ are replaced with their parents (In Step 2, the parent of $v_9$ is $v_{12}$, which is not shown in \reffig{existgpuscan} due to limited space). Since $v_8$ has two neighbors $v_1$ and $v_{12}$ in $E'.v$, it is a hub vertex. $v_3$, $v_5$, $v_6$ only have one neighbor in $E'.v$, they are outlier vertices.
\end{example}



\begin{theorem}
\label{thm:gpuscantc}
Given a graph $G$ and two parameters $\epsilon$ and $\mu$, the work of \gpuscan to finish structural clustering is $ O(\Sigma_{(u, v) \in E(G)} (\mydeg(u)+\mydeg(v))+ c \cdot m \cdot \log m)  $, and the span  is $ O( \mydeg_{\kw{max}} +c \cdot \log^2 m) $, where $ c $ denotes the number of iterations in Phase 2.
\end{theorem}

\myproof In Phase 1, for each edge $(u, v)$, \gpuscan has to explore the neighbors of $u$ and $v$. Thus, the work and span of Phase 1 are $O(\Sigma_{(u, v) \in E(G)} (\mydeg(u)+\mydeg(v)))$ and $O(\mydeg_{\kw{max}})$, respectively. In Phase 2, in each iteration, the time used for \kw{sort()} dominates the whole time of this iteration, and the work and span of \kw{sort()} are $O(m \cdot \log m)$ and $O(\log^2m)$ \cite{singh2018survey}. Thus, the work and span of Phase 2 are $O(c \cdot m \cdot \log m)$ and  $O(c \cdot \log^2 m)$, respectively. In Phase 3, \gpuscan  explores the edges once and only the neighbors of hubs and outliers are visited. Thus, the work and span of Phase 3 are $O(m)$ and $O(1)$. Thus, the  work and space of \gpuscan are $ O(\Sigma_{(u, v) \in E(G)} (\mydeg(u)+\mydeg(v))+ c \cdot m \cdot \log m)  $ and $ O( \mydeg_{\kw{max}} +c \cdot \log^2 m) $.

%
%
%
%

\stitle{Drawbacks of \gpuscan.} \gpuscan seeks to parallelize \kw{SCAN}  to benefit from the computational power provided by GPUs. However, it  suffers from the following two drawbacks, which make it unable to fully utilize  the massive parallelism of GPUs. 

\begin{itemize}[leftmargin=*]
  \item \emph{High extra parallelization cost.} As shown in \cite{Chang2017pSCAN}, the time complexity of the best serial \kw{SCAN} algorithm is  $O(m \cdot \mydeg_{max} )$. On the other hand, \refthm{gpuscantc} shows that the work of \gpuscan is $ O(\Sigma_{(u, v) \in E(G)} (\mydeg(u)+\mydeg(v))+ c \cdot m \cdot \log m) $. Obviously,  lots of extra costs are introduced in \gpuscan for parallelization. 
  \item \emph{Low scalability.} \gpuscan  assumes that the input graphs can be fit in the GPU memory. However, the sizes of real graphs such as social networks and web graphs are huge while the GPU memory is generally in the range of a few gigabytes. The assumption that the sizes of data graphs cannot exceed the GPU memory makes it unscalable to handle large graphs in practice. 
\end{itemize}

\section{Our Approach}
\label{sec:inmemory}

We address the drawbacks of \gpuscan in two sections. In this section, we focus on the scenario in which the graphs can be stored in the GPU memory and aim to reduce the high computation cost introduced for parallelization. In the next section, we present our algorithm to address the problem that the graph cannot fit in the GPU memory.   


\subsection{A CSR-Enhanced Graph Layout}
\label{sec:csrenhanced}



\gpuscan simply uses the edge array and adjacent array to represent the input graph. The shortcomings of this representation are two-fold: (1) the degree of a vertex can not be easily  obtained, which is a commonly used operation during the clustering, (2) the relation between the edge array and adjacent array  for the same edge is not directly maintained, which complicates the processing  when identifying the cluster vertices in Phase 2.  To avoid these problems, we adopt a CSR (Compressed Sparse Row)-enhanced graph layout, which contains four parts:  

\begin{figure}[h]
	\vspace{-0.1cm}
	\centerline{\includegraphics[width=0.49\textwidth]{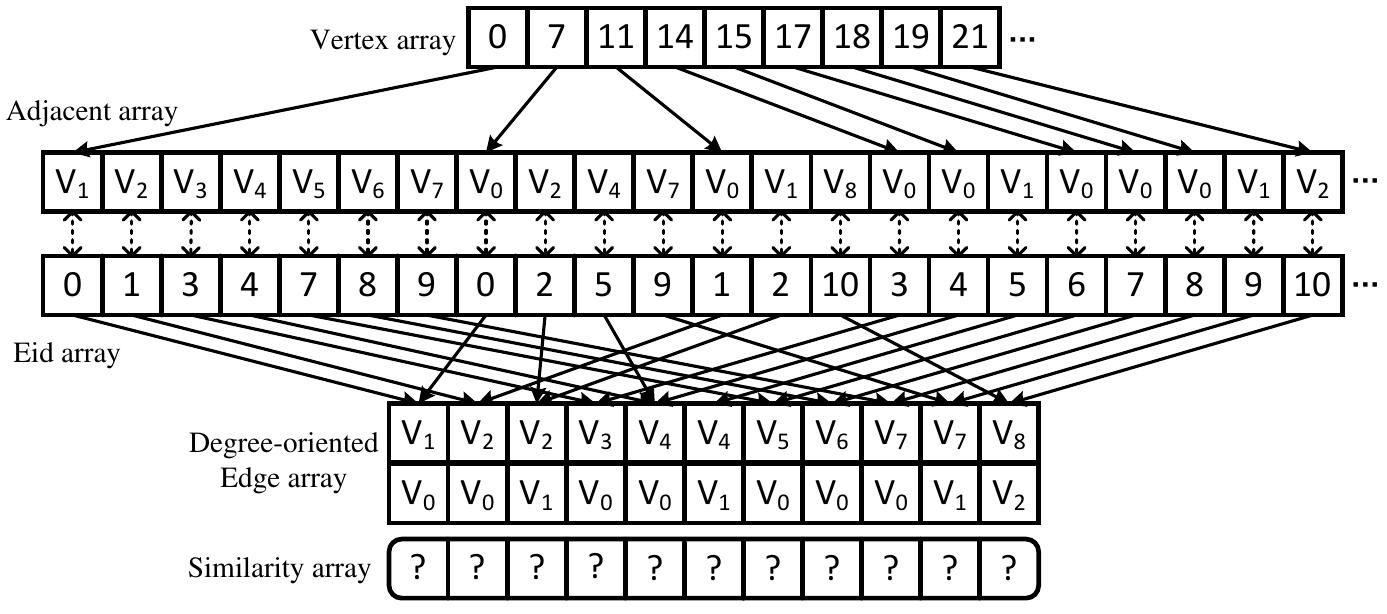}}
	\caption{A CSR-Enhanced graph layout for $ G'$}
	\label{fig:workload}
	\vspace{-0.1cm}
\end{figure}

\begin{itemize}[leftmargin=*]
  \item Vertex array. Each entry keeps the starting index for a specific vertex in the adjacent array.
  \item Adjacent array. The adjacent array contains the adjacent vertices of each vertex. The adjacent vertices are stored in the increasing order of vertex ids.
  \item Eid array. Each entry corresponds to an edge in the adjacent array and keeps the index of the edge in the degree-oriented edge array.
  \item Degree-oriented edge array. The degree-oriented edge array contains the edges of the graph and the edges with the same source vertex are stored together. For each edge $(u, v)$ in the degree-oriented edge array, we guarantee $\mydeg(u) < \mydeg(v)$ or $\mydeg(u)  = \mydeg(v)$ and $\kw{id}(u) < \kw{id}(v)$.  In this way, the workload for edges incident to a large degree vertex can be computed based on the vertex with smaller degree.
  \item Similarity array. Each element corresponds to an edge in the degree-oriented edge array and maintain the similarity of the corresponding edge.
\end{itemize}

\begin{example}
	\reffig{workload} shows the CSR-Enhanced graph layout regarding $v_0$, $v_1$, $\dots$, $v_8$ in $G'$. These five arrays are organized as discussed above. 
\end{example}

\subsection{A Progressive Structural Graph Clustering Approach}

To reduce the high parallelization cost,  the designed algorithm should avoid unnecessary computation during the clustering and fully exploit the unique characteristics of GPU structure. As shown in \refsec{gpuscan}, \gpuscan first computes the structural similarity for each edge in Phase 1 and then identifies the roles of each vertex accordingly. However, based on \refthm{gpuscantc}, computing the structural similarity is a costly operation. To avoid unnecessary structural similarity computation, we adopt a progressive approach in which the lower bound ($\underline{N_{\epsilon}}[v]$) and upper bound ($\overline{N_{\epsilon}}[v]$) of $N_{\epsilon}[v]$ are maintained. Clearly, we have the following lemma on $\underline{N_{\epsilon}}[v]$ and  $\overline{N_{\epsilon}}[v]$:

\setlength{\textfloatsep}{2pt}
\SetInd{0.4em}{0.4em}
\begin{algorithm}[h]
	\DontPrintSemicolon
	\small{
		\caption{$\kw{\gpuscanp}(G, \mu, \epsilon)$}
		\label{alg:gpuscanp}
		
		\For{$v \in V$ {\bf in  parallel}}
		{
			$ \underline{N_{\epsilon}}[v]\gets 0$; $\overline{N_{\epsilon}}[v] \gets \kw{VA}[v+1] - \kw{VA}[v]$; $\kw{role}[v] \gets \large\textcircled{\footnotesize{\bf{?}}}$;
		}
		
		\For{$(u, v) \in E$ {\bf in  parallel}}
		{
			$\kw{sim}(u, v) \gets \large\textcircled{\footnotesize{\bf{?}}}$;
		}
		
		$\kw{identifyCore(G, \mu, \epsilon)}$\;
		$\kw{detectClusters(G, \epsilon)}$\;
		$\kw{classifyHubOutlier}(G)$\;
	}
\end{algorithm}

\begin{lemma}
\label{lem:bound}
Given a graph $G$ and two parameters $\mu$ and $\epsilon$, for a vertex $v \in V(G)$, if $\underline{N_{\epsilon}}[v] \geq \mu$, then $v$ is a core vertex. If $\overline{N_{\epsilon}}[v] \leq \mu$, then $v$ is not a core vertex.
\end{lemma}

According to \reflem{bound},  we can maintain the $\underline{N_{\epsilon}}[v]$ and $\overline{N_{\epsilon}}[v]$ progressively and delay the structural similarity computation until necessary. Following this idea, our new algorithm, \gpuscanp, is shown in \refalg{gpuscanp}. It first initializes   $\underline{N_{\epsilon}}[v]$ as $1$, $\overline{N_{\epsilon}}[v]$ as $\mydeg(v)+1$ for each vertex $v$ (lines 1-2,  \kw{VA} refers to the vertex array). The role of each vertex and the structural similarity of each edge are initialized as \emph{unknown} ($\large\textcircled{\footnotesize{\bf{?}}}$) (lines 1-4). Then, it identifies the core vertices (line 5), detects the clusters follow the core vertices (line 6), and classifies the hub vertices and outlier vertices (line 7).

\stitle{\underline{Identify core vertices.}} \gpuscanp first identifies the core vertices in the graph based on the given parameters $\mu$ and $\epsilon$, which is shown in \refalg{identifycore}.  


\setlength{\textfloatsep}{2pt}
\SetInd{0.4em}{0.4em}
\begin{algorithm}
\DontPrintSemicolon
\small{
	\caption{$\kw{identifyCore}(G, \mu, \epsilon)$}
	\label{alg:identifycore}
	\SetKwFunction{Intersection}{Intersection}
	\For{$(u, v) \in E(G)$ {\bf in warp-level parallel}}{
	\vspace{0.2em}
		\If{$ \kw{role}[u]= \large\textcircled{\footnotesize{\bf{?}}} \vee \kw{role}[v]=\large\textcircled{\footnotesize{\bf ?}}$}{
			$\kw{isSim}\gets \kw{checkSim}(u, v, \epsilon)$;\;
			// the first thread in the warp\;
			\If{$\kw{isSim} =$ {\bf true}}{
				$\kw{sim}(u,v)\leftarrow \large\textcircled{\footnotesize{\bf S}} $;\;
				$\kw{atomicAdd}(\kw{\underline{N_{\epsilon}}}[u],1)$;$\kw{atomicAdd}(\kw{\underline{N_{\epsilon}}}[v],1)$;\;
				\textbf{if} $ {\underline{N_{\epsilon}}}[u]\geq\mu $ \textbf{then} $ \kw{role}[u]\leftarrow \large\textcircled{\footnotesize{\bf C}} $;\;
				\vspace{0.2em}
				\textbf{if} $ {\underline{N_{\epsilon}}}[v]\geq\mu $ \textbf{then} $ \kw{role}[v]\leftarrow \large\textcircled{\footnotesize{\bf C}} $;
			}
			\Else{
				$ \kw{sim}(u,v)\leftarrow  \large\textcircled{\footnotesize{{\bf !S}}}$;\;
				$\kw{atomicAdd}(\kw{\overline{N_{\epsilon}}}[u],-1)$;$\kw{atomicAdd}(\kw{\overline{N_{\epsilon}}}[v],-1)$;\;
				\textbf{if} $ {\overline{N_{\epsilon}}}[u]<\mu $ \textbf{then} $ \kw{role}[u]\leftarrow \large\textcircled{\footnotesize{\bf{!C}}} $;\;
				\vspace{0.2em}
				\textbf{if} $ {\overline{N_{\epsilon}}}[v]<\mu $ \textbf{then} $ \kw{role}[v]\leftarrow  \large\textcircled{\footnotesize{\bf !C}}$;
			}
		}
	}}
	\vspace{0.1cm}
	
	\SetKwProg{myproc}{Procedure}{}{}
	\myproc{{$\kwnospace{checkSim}(u, v, \epsilon)$ }}{
	$\kw{sum} \gets 0$;\;
	\For{$w\in \nbr(u)$ \bf{in}  \bf{parallel}}{	
		$ \kw{low} \leftarrow \kw{VA}[v];\;\kw{high}\leftarrow \kw{VA}[v+1] $;\;
		
		\While{$\kw{low}<\kw{high} $}{
			$ \kw{mid}\leftarrow (\kw{low}+\kw{high})/2 $;\;
			\uIf{$ w<\kw{AdjA}[\kw{mid}] $}{$ \kw{high}\leftarrow \kw{mid}; $\;}
			\ElseIf{$ w>\kw{AdjA}[\kw{mid}] $}{$ \kw{low}\leftarrow \kw{mid}+1; $\;}
			\Else{$\kw{atomicAdd}(\kw{sum}, 1) $; \bf{break};\;}
		}
		
	}
	$ d_u\leftarrow \kw{VA}\lbrack u+1\rbrack-\kw{VA}\lbrack u\rbrack$;\;
		$ d_v\leftarrow \kw{VA}\lbrack v+1\rbrack-\kw{VA}\lbrack v\rbrack$;\;
	\bf{return} $ \kw{sum}+2\geq\epsilon\cdot\sqrt{(d_u+1)\cdot (d_v+1)} $;\;}
\end{algorithm}

To identify the core vertices, it assigns a warp for each edge $(u, v)$ to determine its structural similarity (line 1), which is similar to \kw{GPUSCAN}. However, the detailed procedures to compute the similarity are significantly different from that of \kw{GPUSCAN}. Specifically,  if the role of $u$ or $v$ has been known, it delays the computation of structural similarity between $u$ and $v$ to the following steps as the role of these two vertices has been determined and it is possible that we can obtain the clustering results without knowing the structural similarity of $u$ and $v$ in the following steps (line 2). Then, it computes the structural similarity between $u$ and $v$ by invoking \kw{checkSim}. If $u$ and $v$ are  similar, then the similarity indicator between $u$ and $v$ is set to $\large\textcircled{\footnotesize{\bf S}}$ (line 6) and  ${\underline{N_{\epsilon}}[u]}$/${\underline{N_{\epsilon}}[v]}$ is increased by $1$ as $u$ and $v$ are similar (line 7). After that, if ${\underline{N_{\epsilon}}[u]}$/${\underline{N_{\epsilon}}[v]} \geq \mu$, the role of $u$/$v$ is set as $\large\textcircled{\footnotesize{\bf C}} $ (lines 8-9). If $u$ and $v$ are  dissimilar, ${\overline{N_{\epsilon}}[u]}$/${\overline{N_{\epsilon}}[v]}$ and the role of $u$/$v$ are set  accordingly (lines 11-14).

\begin{figure}[t]
	\vspace{-0.2cm}
	\centering
	\subfigure[Find neighbor of $ v_0 $ and $ v_1 $]{
		\begin{minipage}[b]{0.33\textwidth}
			\includegraphics[width=1\textwidth]{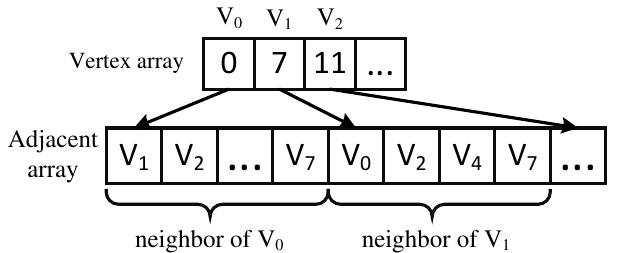}
			\vspace{-0.5cm}
		\end{minipage}
	}
	\vspace{-0.2cm}
	\\
	\subfigure[One thread per vertex in $\nbr(v_1) $]{
		\begin{minipage}[b]{0.33\textwidth}
			\includegraphics[width=1\textwidth]{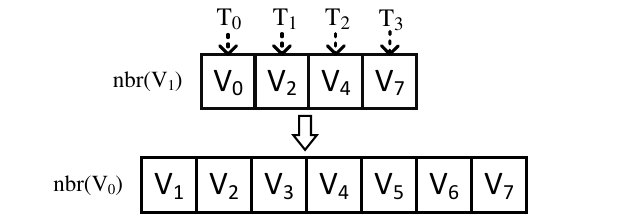}
			\vspace{-0.5cm}
		\end{minipage}
	}
	\vspace{-0.2cm}
	\\
	\subfigure[Parallel binary search in $\nbr(v_0) $]{
		\begin{minipage}[b]{0.33\textwidth}
			\includegraphics[width=1\textwidth]{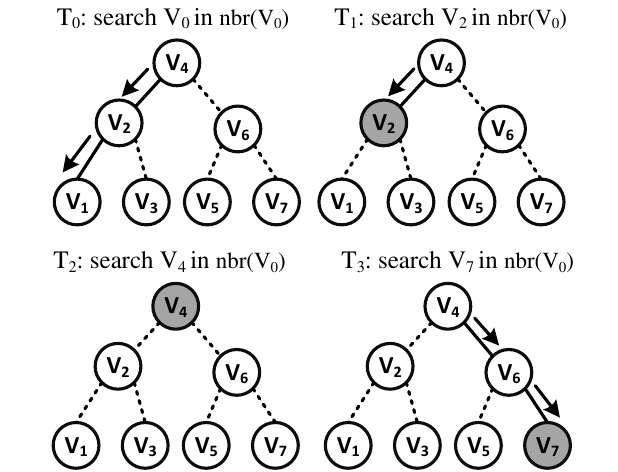}
			\vspace{-0.5cm}
		\end{minipage}
	}
	\vspace{-0.2cm}
	\caption{Check Similarity between $v_0$ and $v_1$ } \label{fig:computation}
\end{figure}

\vspace{-0.02cm}

Procedure \kw{checkSim} is used to determine the structural similarity between $u$ and $v$ based on the given $\epsilon$. Without loss of generality, we assume $\mydeg(u) < \mydeg(v)$. For each neighbor $w$ of $u$, it assigns a thread to check whether $w$ is also a neighbor of $v$ (line 17). Since the neighbors of $v$ are stored in increasing order based on their id in \kw{AdjA}, it looks up $w$ among the neighbors of $v$ in a binary search manner (lines 18-26). Specifically, three indices \kw{low}, \kw{high}, and \kw{mid} are maintained and initialized as $\kw{VA}[v]$, $\kw{VA[v+1]}$, and $(\kw{low}+\kw{high})/2$, respectively. If $w$ is less than $\kw{AdjA}[\kw{mid}]$,  \kw{high} is updated as \kw{mid} (lines 21-22) (\kw{AdjA} refers to the adjacent array).  If $w$ is larger than $\kw{AdjA}[\kw{mid}]$, then \kw{low} is updated as \kw{mid}+1 (lines 23-24). Otherwise, the common neighbors of $u$ and $v$, which are recorded in \kw{sum}, are increased by 1 (line 26). Last, whether $u$ and $v$ are similar returns based on \refdef{ss} (line 27-29).

\begin{example}
\reffig{computation} shows the procedures of \kw{checkSim} to check the similarity between $v_0$ and $v_1$ of $G'$ shown in \reffig{graph}. In order to compute their similarity, the neighbors of $ v_1 $ and $ v_0 $ are first obtained from the CSR-Enhanced graph layout as shown in \reffig{computation} (a).  As the $\mydeg(v_1) < \mydeg(v_0)$, thread $ T_0 $, $ T_1 $, $ T_2 $, and $ T_3 $ in the warp are used to check whether the neighbors $v_0, v_2, v_4, v_7$ of $v_1$ are also neighbors of $v_0$, which is shown in \reffig{computation} (b). Take thread $T_0$ as an example, since the neighbors of $v_0$ are sorted based on their ids, thread $T_0$ explores the neighbors of $v_0$ in a binary-search manner, which is shown in \reffig{computation} (c). After traversing the leaf node $v_1$, it is clear that $v_0$ is not a neighbor of $v_0$. Similarly, threads $T_1-T_3$ find that $v_2$, $v_4$, and $v_7$ are neighbors of $v_0$ as well. Therefore, the common neighbors of $v_0$ and $v_1$ are $v_2$, $v_4$, $v_7$. Because $2+3 > 0.6 \times \sqrt{5\times8}$, $v_0$ and $v_1$ are similar, and \kw{checkSim} returns \kw{true}.
\end{example}
%


\setlength{\textfloatsep}{2pt}
\SetInd{0.4em}{0.4em}
\begin{algorithm}
	\caption{$\kw{detectClusters}(G, \epsilon)$}
	\label{alg:detectclusters}
	\DontPrintSemicolon
	\For{$ u\in V $ {\bf in  parallel}}{
		{\bf if} $\kw{role}[u] = \large\textcircled{\footnotesize{\bf C}} $ {\bf then} $ \kw{parent}[u]\leftarrow u $; $ u.\kw{height}\leftarrow1 $;\; {\bf else} $\kw{parent}[u]\leftarrow -2$; \;
	}
	\For{$ u\in V \wedge \kw{role}[u]={\large\textcircled{\footnotesize{\bf C}}} $  {\bf in warp-level parallel}}{
		$ u_p \leftarrow $ $\kw{root}(u, \kw{parent})$;\;
		\For{$v \in \nbr(u)$ {\bf in  parallel} }{
			\If{$\kw{role}[v]= \large\textcircled{\footnotesize{\bf C}} \wedge \kw{sim}(u,v)= \large\textcircled{\footnotesize{\bf S}} $}{
				$ v_p\leftarrow \kw{root}( v ,  \kw{parent} )$;\;
				{\bf if}{ $ u_p \neq v_p $} {\bf{then}} {$\kw{union}(u_p ,  v_p ,  \kw{parent})$;\;}
			}
		}
	}
	\For{$ u\in V \wedge \kw{role}[u]= \large\textcircled{\footnotesize{\bf C}}$ {\bf in warp-level parallel}}{
		$ u_p\leftarrow  \kw{root}( u ,  \kw{parent}) $; \;
		\For{$v \in \nbr(u)$ \bf{in warp-level parallel}}{
			\If{$ \kw{role}[v]=\large\textcircled{\footnotesize{\bf C}} \wedge \kw{sim}(u,v)=\large\textcircled{\footnotesize{\bf ?}}  $}{
				$ v_p \leftarrow \kw{root}( v ,  \kw{parent} )$;\;
				\If{$ u_p \neq v_p $}{
					\If{$ \kw{checkSim}(u, v, \epsilon) $}{ $\kw{sim}(u,v) \gets \large\textcircled{\footnotesize{\bf S}}$; $\kw{union}(u_p , v_p ,  \kw{parent}) $; \;}
					{\bf{else}}{ $ \kw{sim}(u,v) \gets \large\textcircled{\footnotesize{\bf !S}}$;}
				}
			}
		}
	}
	\For{$  u\in V \wedge \kw{role}[u]= \large\textcircled{\footnotesize{\bf C}} $ \bf{in parallel}}{
		$ \kw{parent}[u] \leftarrow \kw{root}( u ,  \kw{parent} )$;\;
	}
	\For{$ u\in V \wedge \kw{role}[u]= \large\textcircled{\footnotesize{\bf 
				C}}$ {\bf in warp-level parallel}}{
		\For{$v \in \nbr(u)$ \bf{in warp-level parallel}}{
			\If{$ \kw{role}[v]=\large\textcircled{\footnotesize{\bf !C}} $}{
				\If{ $ \kw{sim}(u,v) = \large\textcircled{\footnotesize{\bf ?}}$}{
					$\kw{sim}(u,v) \gets \kw{checkSim}(u, v, \epsilon)?\, \large\textcircled{\footnotesize{\bf S}}:\large\textcircled{\footnotesize{\bf !S}}$;
				}
				\If{$\kw{sim}(u,v)=\large\textcircled{\footnotesize{\bf S}}  $}{
					$ \kw{parent}[v]\gets \kw{parent}[u] $;\;
					
				}
			}
			
		}
	}
\vspace{0.1em}
\SetKwProg{myproc}{Procedure}{}{}	
\myproc{{$\kwnospace{root}(u, \kw{parent})$ }}{
	$ \kw{p_u}\leftarrow \kw{parent}[u] $; $\kw{next}\leftarrow \kw{parent}[\kw{p_u}]$;\;
%
	\While{$\kw{p_u} \neq \kw{next}$}{
		$ \kw{p_u}\leftarrow \kw{next} $;
		$ \kw{next}\leftarrow\kw{parent}[\kw{p_u}] $;\;}
	{\bf return} $ \kw{p_u} $\;}
\vspace{0.1em}
\SetKwRepeat{Do}{do}{while}
\myproc{{$\kwnospace{union}(u, v, \kw{parent})$ }}{	
	{\bf if} $u.\kw{height} < v.\kw{height}$ {\bf then} $\kw{swap}(u, v)$;\;
	$ \kw{fg}\leftarrow$ {\bf true}; \;
	\While{\kw{fg}}{
		$ \kw{fg}\leftarrow${\bf false}; $ \kw{res}\;\leftarrow\;\kw{atomicCAS}(\kw{\&parent}[v],v,u) $;\;
				{\bf if} {$ \kw{res} \neq v $} {\bf then}{
					$ v\leftarrow \kw{res};\kw{fg}\leftarrow$ {\bf true}; }\;		
			}
		{\bf if} {$ u \neq v \wedge u.\kw{height}=v.\kw{height}$} {\bf then}{
			$\kw{atomicADD}(u.\kw{height},1) $;}
		}

\end{algorithm}

\stitle{\underline{Detect clusters.}} After  the core vertices have been identified,  \gpuscanp detects clusters in the graph, which is shown in \refalg{detectclusters}. After the core vertices identification phase, the vertices are either core vertices or non-core vertices. Moreover, the similarities of some edges are known but there are still edges whose similarities are unknown. Therefore, \refalg{detectclusters} first detects the sub-clusters that can be determined by the computed information (lines 1-9). After that, it further computes the similarity for the edges whose similarities are unknown and obtains the complete clusters in the graph (lines 10-27). In this way, we can avoid some unnecessary similarity computation compared with \kw{GPUSCAN}.


Specifically, \refalg{detectclusters} uses the \kw{parent} array to store the cluster id  that each vertex belongs to (negative cluster id value indicates the corresponding vertex is not in a cluster currently). Similar to \kw{GPUSCAN}, \refalg{detectclusters} also uses a spanning forest to represent the clusters through the \kw{parent} array. However, \refalg{detectclusters} uses a different algorithm to construct the forest, which makes \gpuscanp much more efficient than \kw{GPUSCAN} as verified in our experiment. It initializes each core vertex $u$ as a separate cluster with cluster id $u$ and the non-core vertices with a negative cluster id (lines 1-3). Then, for each core vertex $u$, it first finds the root id of the cluster by the procedure \kw{root} (line 5). After that, for the neighbors $v$ of $u$, if we have already known $u$ and $v$ are similar (line 7), we know $u$ and $v$ are the same cluster following \refdef{cluster}. Thus, we union the subtrees represented by $u$ and $v$ in \kw{parent}  through procedure \kw{union} (lines 6-9).      

After having explored the sub-clusters that can be determined by the computed information, for each core vertices $u$, \refalg{detectclusters} further explores the neighbors $v$ of $u$ such that $v$ is also a core vertex but the similarity of $u$ and $v$ are unknown (lines 10-13). If $u$ and $v$ are not in the same cluster currently (line 15), then it computes the similarity between $u$ and $v$ (line 16). If $u$ and $v$ are similar, we  union the subtrees represented by $u$ and $v$ in the \kw{parent} array (line 17). Otherwise, $u$ and $v$ are marked as dissimilar (line 18). When the clusters of the core vertices are determined, the root cluster id is assigned to each core vertex (lines 19-20). At last, \refalg{detectclusters} explores the non-core neighbors $v$ of each core vertex $u$, if the similarity between $u$ and $v$ is unknown, their similarity is computed (lines 24-25). If $u$ and $v$ are similar, $v$ is merged in the cluster represented by $u$ (lines 26-27). For the procedure \kw{root} and \kw{union}, they are used to find the root of the tree in the forest and union two sub-trees in the forest. As they are self-explainable, we omit the description. Note the  $u.\kw{height}$/$v.\kw{height}$ in line 34 represents the height of tree rooted at $u/v$.

\begin{figure}[htbp]
	\vspace{-0.2cm}
	\centering
	\subfigure[Process core vertices $ v_1$  and $ v_4$]{
		\begin{minipage}[b]{0.22\textwidth}
			\includegraphics[width=1\textwidth]{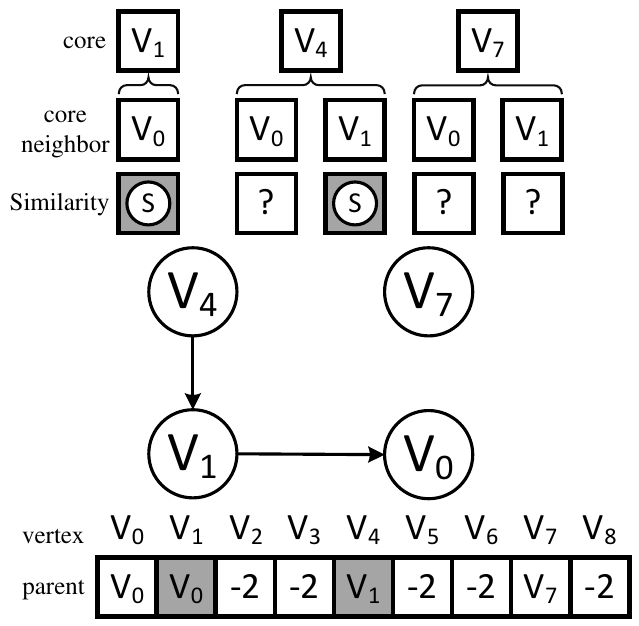}
			\vspace{-0.6cm}
		\end{minipage}
	}
	\subfigure[Process $v_4$ and $v_0$]{
		\begin{minipage}[b]{0.22\textwidth}
			\includegraphics[width=1\textwidth]{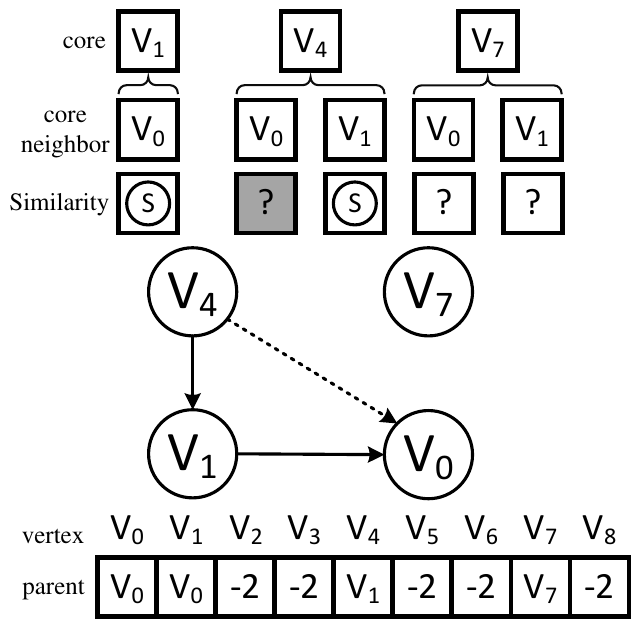}
			\vspace{-0.6cm}
		\end{minipage}
	}
	\vspace{-0.2cm}
	\\
	\subfigure[Process $ v_7$ and $v_0$]{
		\begin{minipage}[b]{0.22\textwidth}
			\includegraphics[width=1\textwidth]{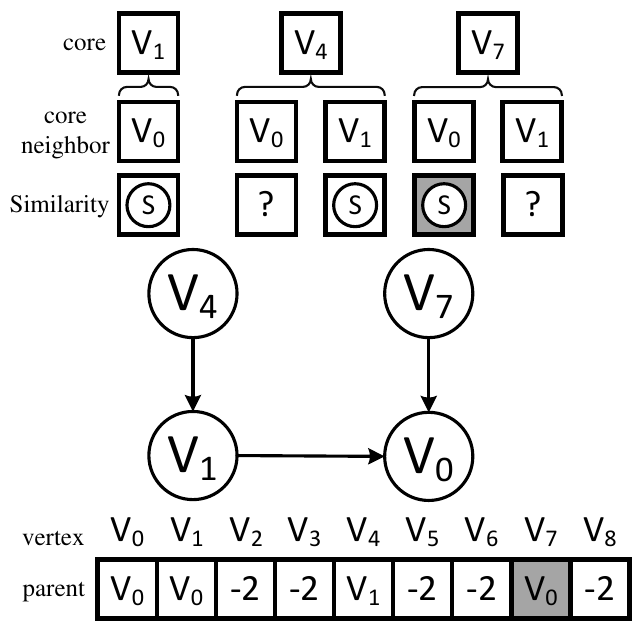}
			\vspace{-0.6cm}
		\end{minipage}
	}
	\subfigure[Process $ v_7$ and $v_1 $]{
		\begin{minipage}[b]{0.22\textwidth}
			\includegraphics[width=1\textwidth]{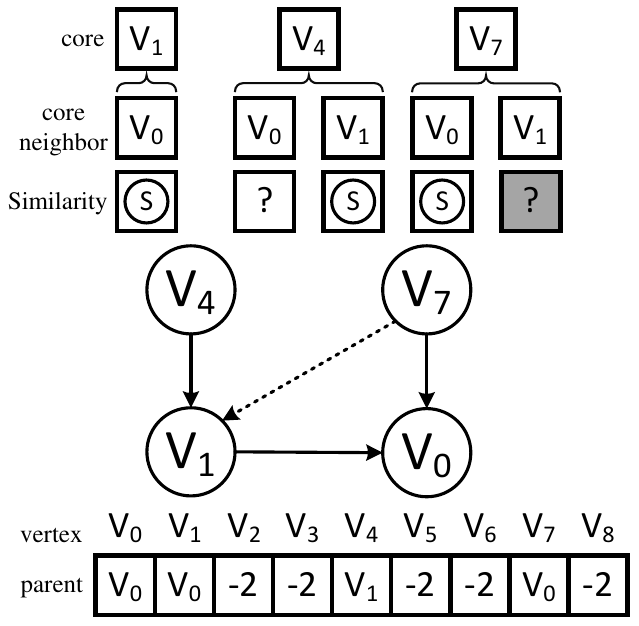}
			\vspace{-0.6cm}
		\end{minipage}
	}
	\vspace{-0.2cm}
	\\
	\subfigure[Trace root vertex]{
		\begin{minipage}[b]{0.22\textwidth}
			\includegraphics[width=1\textwidth]{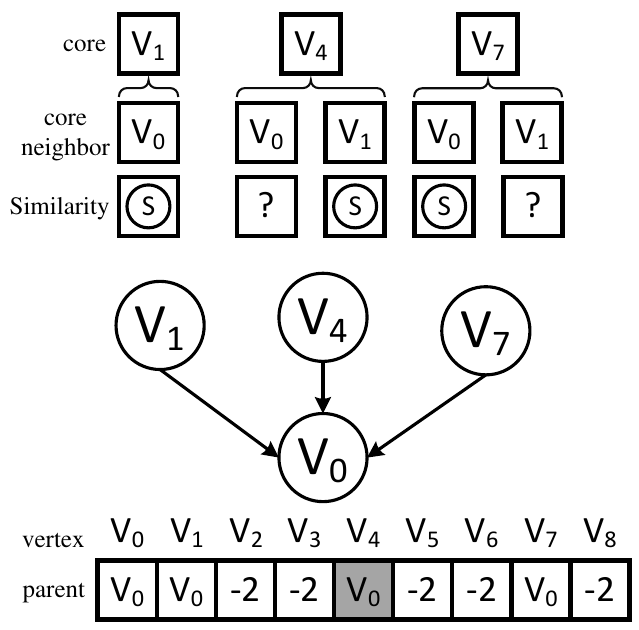}
			\vspace{-0.6cm}
		\end{minipage}
	}
	\subfigure[Cluster non-core $ v_2 $]{
		\begin{minipage}[b]{0.22\textwidth}
			\includegraphics[width=1\textwidth]{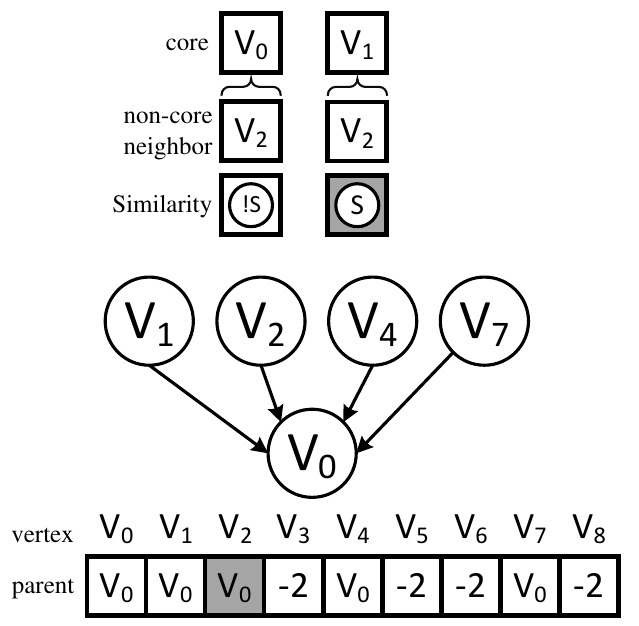}
			\vspace{-0.6cm}
		\end{minipage}
	}
	\vspace{-0.2cm}
	\\
	\caption{Detect clusters in $G'$}
	\label{fig:clustering}
\end{figure}

\begin{example}
	\reffig{clustering} shows the procedure to detect clusters  in $G'$. 
	Assume that we know that $v_0$, $v_1$, $v_4$, $v_7$ are core vertices, $(v_1, v_0) $ and $ (v_4, v_1) $ are similar, but the similarity of $ (v_4, v_0) $, $ (v_7, v_0) $ and $ (v_7, v_1) $ are unknown. Parents of the four core vertices are initialized as themselves while the others are set as -2. 
	
	\gpuscanp first processes the core vertices in parallel. To avoid redundant computation, when clustering core vertices, only neighbors  with  smaller ids than their own ids are considered. At the same moment, for $v_1$, $ (v_1, v_0) $ are similar, then the parent of $v_1$ becomes $v_0$ and for $v_4$, $(v_4, v_1)$ are similar, then the parent of $v_4$ becomes $v_1$, which are shown in \reffig{clustering} (a). After that, \gpuscanp detects the clusters related to $v_4$ and $v_7$ as the similarities between some of their neighbors and themself have not been determined yet. For the neighbor $v_0$ of $v_4$, \gpuscanp finds that $v_0$ and $v_4$ have been in the same cluster by  procedure \kw{root}, they do not need further process, which is shown in \reffig{clustering} (b). For $v_7$, due to the atomic operation, the thread which explores its neighbor $v_0$ executes first. It finds that they are not in the same cluster based on procedure \kw{root}. Therefore, after determining that they are similar by procedure \kw{checkSim}, it changes the parent of $ v_7 $ to $ v_0 $, which is shown in \reffig{clustering} (c). Then, $v_7$ and $v_1$ are processed by another thread similarly, which is shown in \reffig{clustering} (d). When the cluster of core vertices $ v_0$, $v_1$, $v_4$, and $v_7$ is detected, \gpuscanp explores these vertices and sets their corresponding element in the \kw{parent} array as the root vertices in the forest. \reffig{clustering} (e) illustrates this process, in which the \kw{parent} of $ v_4 $ changes to the traced root vertex $ v_0 $. At last, \gpuscanp assigns the parent of $ v_1$ to $ v_2 $ based on the similarity of core vertex $ v_1 $ and non-core vertex $ v_2 $ as shown in \reffig{clustering} (f).
	\end{example}

\begin{algorithm}[t]
	\caption{$\kw{classifyHubOutlier}(G)$}
	\label{alg:classifyhuboutlier}
	\DontPrintSemicolon
		\For{$u\in V\wedge \kw{parent}[u]<0 $ {\bf in warp-level parallel}}{
			$\kw{parent}[u]\leftarrow -2$; $ \kw{role}[u]= \large\textcircled{\footnotesize{\bf O}} $;\;
			\If{$ |\nbr(u)| > 1$}{
				$ \kw{n_{c}}\leftarrow -1$;\;
				\For{$v \in \nbr(u)$ {\bf in  parallel} }{
					$\kw{id} \leftarrow -1$;\;
					\If{$ \kw{parent}[v]>=0 $}{
						$ \kw{id}\leftarrow\kw{atomicCAS}(\kw{\&n_c},-1,\kw{parent}[v]) $;\;
						\If{$\kw{id}>=0\wedge\kw{id}\ne\kw{parent}[v]$}{
							$\kw{parent}[u]\leftarrow -1$; $ \kw{role}[u]= \large\textcircled{\footnotesize{\bf H}} $;\;
							$ \kw{break} $; \;
						}
					}
				}
			}
	    }
		
\end{algorithm}

\stitle{\underline{Classify hub and outlier vertices}.} At last, \gpuscanp  classifies vertices that are not in clusters into outliers and hubs, which is shown in \refalg{classifyhuboutlier}. Specifically, it assigns a warp for each vertex $u$  not in a cluster and considers $u$ as an outlier (line 2). Then it checks whether $u$ has more than one neighbor vertices, and if this condition is satisfies, $u$ has the potential to become hubs (line 3). Then, \refalg{classifyhuboutlier} uses the shared variable $\kw{n_c}$ of the threads in a warp to record the cluster that one of its neighbors belongs to and sets it as $-1$ (line 4). Each thread in the warp maintains an exclusive variable $\kw{id}$ with an initial value of $ -1 $ (line 5). If a neighbor $v$ of $u$ is found in a cluster by a thread, then, $ \kw{n_c}$ is updated to the cluster id of $v$ by \kw{atomicCAS} (lines 7-8). Once another thread finds that the cluster id recorded by $ \kw{n_c}$ is different from the cluster id of $v$ it processes, it means at least two neighbors of $u$ are in different clusters, therefore, $u$ is classified as a hub (lines 9-11).

\subsection{Theoretical Analysis}

\begin{theorem}
	\label{thm:gpuscanptc}
	Given a graph $G$ and two parameters $\epsilon$ and $\mu$, the work of \gpuscanp finishes the clustering in $O(\Sigma_{(u, v) \in E(G)}\mydeg(u) \cdot \log \mydeg(v))$, and the span is $O(\log \mydeg_{\kw{max}} + \log n)$. 
\end{theorem}

\myproof  In Phase 1 (\refalg{identifycore}), the work and span of \gpuscanp are $O(\Sigma_{(u, v) \in E(G)}\mydeg(u) \cdot \log \mydeg(v))$ and $O(\log \mydeg_{max})$. In Phase 2 (\refalg{detectclusters}), \gpuscanp is possible to check the similarity of all edges, thus, the work of Phase 2 is $O(\Sigma_{(u, v) \in E(G)}\mydeg(u) \cdot \log \mydeg(v))$ as well. For the span, the height of the spanning forest is $O(\log n)$ due to the union-by-size used in line 37 of \refalg{detectclusters} \cite{simsiri2018work}. Thus, the span of Phase 2 is $O(\log n)$. In Phase 3 (\refalg{classifyhuboutlier}), it is easy to derive that the work is $O(m)$ and $O(1)$. Therefore, the work of \gpuscanp is $O(\Sigma_{(u, v) \in E(G)}\mydeg(u) \cdot \log \mydeg(v))$ and the span is $O(\log \mydeg_{\kw{max}} + \log n)$.

Compared with \gpuscan, \refthm{gpuscanptc} shows that \gpuscanp reduces the total work and the span during the clustering. Note that in Phase 1, \gpuscanp uses the binary-search-based manner the check the similarity, which reduces the span of Phase 1 from $O(\mydeg_{\kw{max}})$ to $O(\log \mydeg_{\kw{max}})$ compared with \gpuscan.

\section{A New Out-of-Core Algorithm}
\label{sec:outofcore}




In the previous section, we focus on improving the clustering performance when the graph can be fit in the GPU memory. However, in real applications, the graph data can be very large and the GPU memory is insufficient to load the whole graph data. A straightforward solution is to use the Unified Virtual Memory (UVM) provided by the GPUs directly. However, this is approach is inefficient as verified in our experiments. On the other hand, in most real graphs, such as social networks and web graphs, the number of edges is much larger than the number of vertices. For example, the largest dataset in SNAP contains 65 million nodes and 1.8 billion edges. It is practical to assume that the vertices of the graph can be loaded in the GPU memory while the edges are stored in the CPU memory.

Following this assumption, we propose an out-of-core GPU \kw{SCAN} algorithm that adopts a divide-and-conquer strategy. Specifically, we first divide the graph into small subgraphs whose size does not exceed the GPU memory size. Then, we conduct the clustering  based on the divided subgraphs instead of the original graph. As the divided subgraphs are much smaller than the original graphs, the GPU memory requirement is significantly reduced. Before introducing our algorithm,  we first define the local subgraphs as follows:   

\begin{definition}
\label{def:localgraph}
\textbf{(Edge Extended Subgraph)} Given a graph $G$ and a set of edges $S \in E(G)$, the edge extended subgraph of $S$ consists of the edges  $S$ and $(u, v) \in E(G)$ such that $(u, v)$ incident to at least one edge in $S$.	
\end{definition}


\begin{lemma}
	\label{lem:divide}
	Given a graph $G$ and a set of edge $S \in E(G)$, let $G_s$ denote the edge extended subgraph of  $S$, then, for any edge $(u, v)\in S$, $\sigma_{G_s}(u, v) = \sigma(u, v)$, where $\sigma_{G_s}(u, v)$ denotes the structural  similarity computed based on  $G_s$.
\end{lemma}




\begin{figure}[htp]
	\vspace{-0.2cm}
	\centerline{\includegraphics[width=0.25\textwidth]{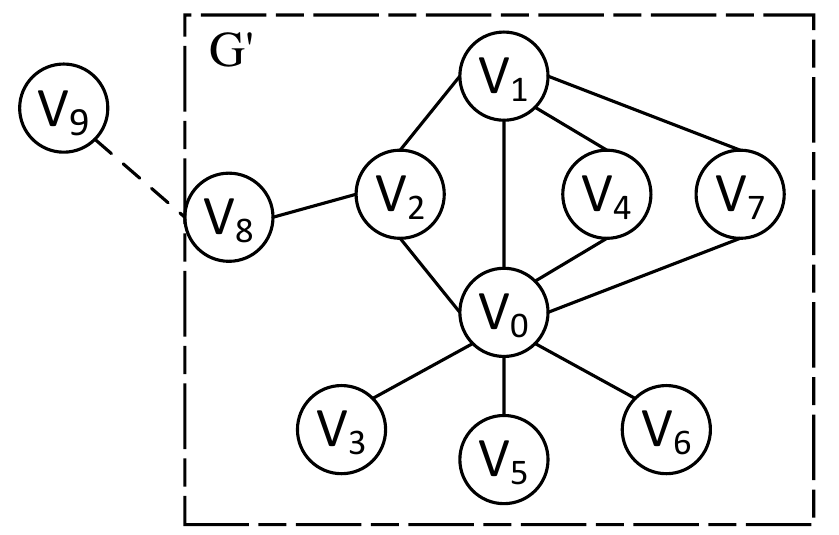}}
	\caption{Edge Extended Subgraph of $G'$}
	\label{fig:extendGraph}
	\vspace{-0.2cm}
\end{figure}

\begin{example}
	\reffig{extendGraph} shows the edge extended subgraphs of $G'$. As $(v_8, v_9)$ and $(v_2, v_8)$ share the same vertex $v_8$ but $(v_8, v_2)$ is not in $G'$, $(v_8, v_2)$ should be added into the edge extended subgraph of $G'$.   
\end{example}

Therefore,  we divide  $G$ into a series of edge extended subgraphs $G_{S_1}, G_{S_2}, \dots, G_{S_k}$ such that $S_i \cup S_2, \dots, \cup S_k = E(G)$, $S_i \cap S_j = \emptyset$, where $1 \leq i \neq  j \leq k$, and each edge extended subgraph can be loaded in the GPU memory. Following \reflem{divide}, we can obtain the structural similarity for all edges in $S$ regarding $G_s$ correctly. Moreover, as the vertices can be loaded in the memory, which means we can identify the core vertices by iterating each edge extended subgraphs and maintain the clustering information in the GPU memory accordingly. 

\begin{algorithm}[h]
	\DontPrintSemicolon
	\caption{$\gpuscanpo(G, \mu, \epsilon, M_g)$}
	\label{alg:outofcore}
	$ G_s(V_s,E_s,\kw{sim}_s)\gets{\emptyset};S\gets G_s;  s\gets0$\;
	
	/* Code Executed by CPU */\;
	\For{$(u, v) \in E(G)$ } 
	{
		$ G'_s(V'_s,E'_s,\kw{sim'}_s)\gets{\emptyset} $\;
		\For{$w \in \nbr(u)/\nbr(v)$}
		{
			\If{$w \notin V_s$}
			{
				$V'_s \gets V'_s \cup \{w\}$; \;
			}
			\If{$(u, w) \notin E_s$}
			{
				$E'_s \gets E'_s\cup\{(u, w)\}$; \;
			}
			
		}
		\If{$ \kw{Memof(G_s)}+\kw{{Memof}(G'_s)}+15 |V(G)|> M_g $}{
			$ s\gets s+1;G_s(V_s,E_s,\kw{sim}_s)\gets{\emptyset};S\gets S\cup G_s $\;
		}
		$ V_s\gets V_s\cup V'_s;E_s\gets E_s\cup E'_s$\;
		
		$\kw{sim}(u, v) \gets \large\textcircled{\footnotesize{\bf{?}}}$; $ \kw{sim}_s\gets \kw{sim}_s\cup\left \{\kw{sim}(u,v)\right \}$\;
		
	}
	
	/* Code Executed by GPU */\;
	\For{$v \in V$ {\bf in  parallel}}
	{
		$ \underline{N_{\epsilon}}[v]\gets 1$; $\overline{N_{\epsilon}}[v] \gets \kw{VA}[v+1] - \kw{VA}[v]+1$; $\kw{role}[v] \gets \large\textcircled{\footnotesize{\bf{?}}}$;
		$ \kw{parent}[v]\gets-2 $;
	}
	\For{$G_s \in S$}{
		$\kw{load(G_s)}$;
		$\kw{identifyCore(G_s, \mu, \epsilon)}$;
		$\kw{store(G_s)}$;
	}
	\For{$G_s \in S$}{
		$\kw{load(G_s)}$;
		$\kw{detectClusters(G_s, \epsilon)}$;
		$\kw{store(G_s)}$;
	}
	\For{$G_s \in S$}{
		$\kw{load(G_s)}$;
		$\kw{classifyHubOutlier}(G_s, \epsilon)$;
	}
	
\end{algorithm}

\stitle{Algorithm.} Following the above idea, our algorithm GPU-based out-of-core algorithm is shown in \refalg{outofcore}. It first partitions the edges $E(G)$ of $G$ into a series of disjoint sets and constructs the corresponding edge extended subgraphs that can be fit in the GPU memory based on the GPU memory capacity $M_g$ following \refdef{localgraph} by CPU (line 3-12). After that, we initialize $\underline{N_{\epsilon}}[v]$, $\overline{N_{\epsilon}}[v]$, $\kw{role}$, and $\kw{parent}$ array for each vertex in GPU memory following our assumption. After that, based on \reflem{divide}, the similarity for each edge can be correctly obtained by the edge extended subgraph alone. Accordingly, the clusters and the role of each vertex can also be determined by the edge extended subgraph locally. Therefore, we just load each edge extended subgraph into the GPU memory sequentially and repeat the three phases by invoking \kw{identifyCore}, \kw{detectClusters}, and \kw{classifyHubOutlier} for the in-GPU-memory algorithm (lines 14-21). The correctness of \refalg{outofcore} can be easily obtained based on \reflem{divide} and the correctness of \refalg{gpuscanp}. The only thing that needs to explain is how to estimate the size of the edge extended subgraph in line 10. In \refalg{outofcore}, $G_s$ represents the current edge extended subgraph (line 1).  Whenever a new edge $ (u, v) $ is to be added into $ G_s $ (line 3), the edges incident to $(u, v)$ are computed and stored by $G'_s$. According to the CSR-Enhanced graph layout introduced in \refsec{csrenhanced}, $\kw{Memof}(G_s)=25\times|E_s|+4\times|V_s| $. As adjacent array, Eid array, and degree-oriented edge array consumes $24 \times |E_s|$ bytes together, the similarity array consumes $|E_s|$ byte, and  vertex array  consumes $4 \times |V_s|$ (we use 4-byte integer to store a vertex id). $\kw{Memof}(G'_s)$ can be computed in the same way. Moreover, $\underline{N_{\epsilon}}[v]$ (2 bytes per element, as each element will not exceed the value of the parameter $\mu$ which is generally not very large), $\overline{N_{\epsilon}}[v]$ (4 bytes for each element), $\kw{role}$ (1 byte for each element),  $\kw{parent}$ (4 bytes for each element), and $\kw{height}$ (4 byte for each element) array for each vertex take $15 \times |V(G)|$ together. If three GPU memory consumptions are larger than the GPU memory $M_g$ in line 10, we generate a new empty edge extended subgraph in line 11; Otherwise, $G'_s$ is added into $G_s$ in line 12.

As verified in our experiment, \refalg{outofcore} can finish the \kw{SCAN} clustering on a graph with 1.8 billion edges using less than 2GB GPU memory efficiently.

\section{Experiments}

This section presents our experimental results. The in-memory algorithms are evaluated  on a machine with an Intel Xeon 2.4GHz CPU equipped with 128 GB main memory and an Nvidia Tesla V100 16GB GPU, running Ubuntu 16.04LTS. The out-of-core algorithms are  evaluated on a machine with Nvidia GTX1050 2GB GPU, running Ubuntu 16.04LTS.

\begin{table}[t]
	\caption{Datasets used in Experiments}
	\label{tab:datasets}
	\vspace{-0.2cm}
	\begin{center}
		\small{
			\setlength{\tabcolsep}{1.50mm}{
			\begin{tabular}{c|c|c|c|c|c}
				\hline
				{Datasets} \cg  &  {Name}  \cg       & $n$ \cg & $m$ \cg & $ \overline{d} $ \cg &$c$\cg\\ \hline
				Enwiki-2022 &  \kw{EW}   & 6,492,490        & 159,047,205         & 24.50 & 91 \\
				IndoChina & \kw{CN}  & 7,414,866	      & 194,109,311       & 26.18   & 143 \\
				Hollywood  &  \kw{HW}   & 2,180,759        & 228,985,632	        &105.00  & 35 \\
				Orkut  &  \kw{OR}   & 3,072,626        & 234,370,166        & 76.28  & 28 \\
				Tech-P2P  &  \kw{TP} & 5,792,297       & 295,659,774        & 51.04  & 173 \\
				UK-2002  &  \kw{UK} & 18,520,486       & 298,113,762        & 16.10  & 201 \\
				EU-2015 & \kw{EU}  & 11,264,052       & 386,915,963        & 34.35 & 230 \\ 
				Soc-Twitter & \kw{ST}  & 21,297,773	      & 530,051,090       & 24.89  & 30 \\
				\hline
				\hline
				Twitter-2010 & \kw{TW}  & 41,652,230	      & 1,468,365,182       & 35.25 & - \\
				Friendster & \kw{FR}  & 65,608,366	      & 1,806,067,135       & 27.53  & -\\ 
				\hline
			\end{tabular}
			}
		}
	\end{center}
	\vspace{-0.2cm}
\end{table}

\begin{figure}[htbp]
	\vspace{-0.2cm}
	\centering
	\begin{minipage}[b]{0.25\textwidth}
		\includegraphics[width=1\textwidth]{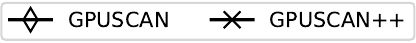}
	\end{minipage}
	\vspace{-0.2cm}
	\\	
	\subfigure[\kw{EW}]{
		\begin{minipage}[b]{0.20\textwidth}
			\includegraphics[width=1\textwidth]{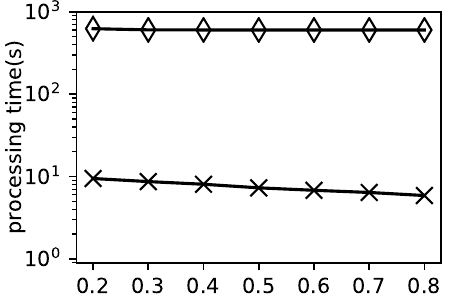}
			\vspace{-0.6cm}
		\end{minipage}
	}
	\subfigure[\kw{CN}]{
		\begin{minipage}[b]{0.20\textwidth}
			\includegraphics[width=1\textwidth]{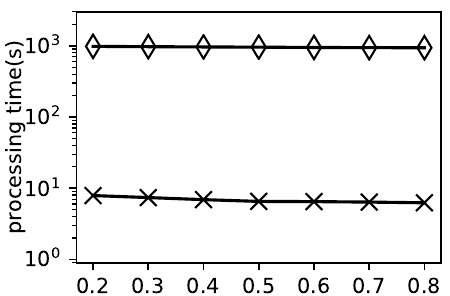}
			\vspace{-0.6cm}
		\end{minipage}
	}
	\vspace{-0.2cm}
	\\
	\subfigure[\kw{HW}]{
		\begin{minipage}[b]{0.20\textwidth}
			\includegraphics[width=1\textwidth]{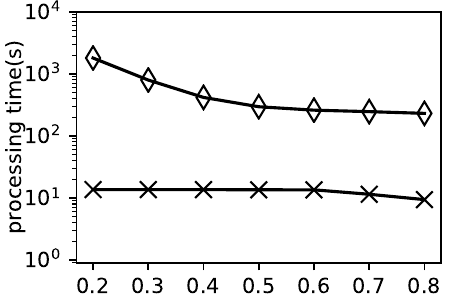}
			\vspace{-0.6cm}
		\end{minipage}
	}
	\subfigure[\kw{OR}]{
		\begin{minipage}[b]{0.20\textwidth}
			\includegraphics[width=1\textwidth]{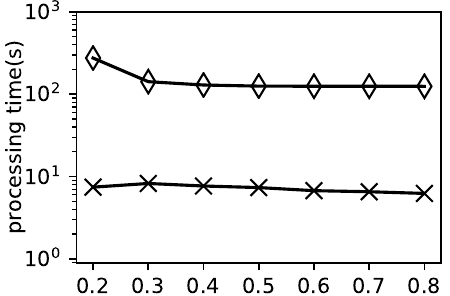}
			\vspace{-0.6cm}
		\end{minipage}
	}
	\vspace{-0.2cm}
	\\
	\subfigure[\kw{TP}]{
		\begin{minipage}[b]{0.20\textwidth}
			\includegraphics[width=1\textwidth]{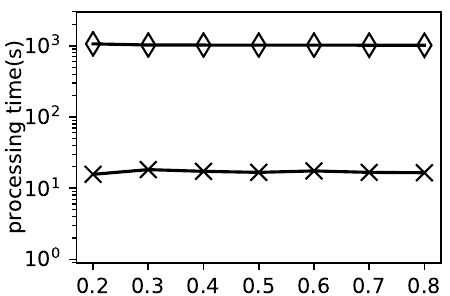}
			\vspace{-0.6cm}
		\end{minipage}
	}
	\subfigure[\kw{UK}]{
		\begin{minipage}[b]{0.20\textwidth}
			\includegraphics[width=1\textwidth]{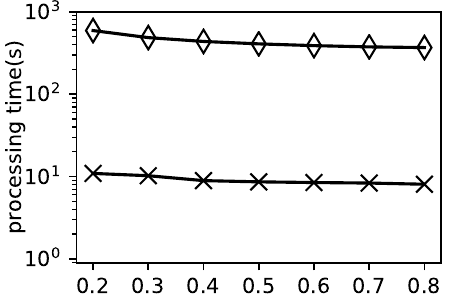}
			\vspace{-0.6cm}
		\end{minipage}
	}
	\vspace{-0.2cm}
	\\
	\subfigure[\kw{EU}]{
		\begin{minipage}[b]{0.20\textwidth}
			\includegraphics[width=1\textwidth]{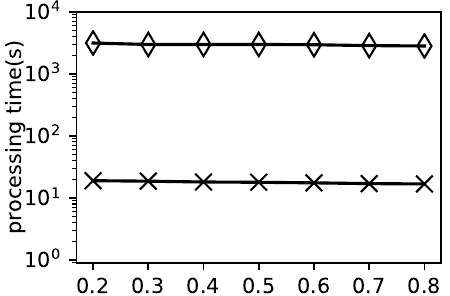}
			\vspace{-0.6cm}
		\end{minipage}
	}
	\subfigure[\kw{ST}]{
		\begin{minipage}[b]{0.20\textwidth}
			\includegraphics[width=1\textwidth]{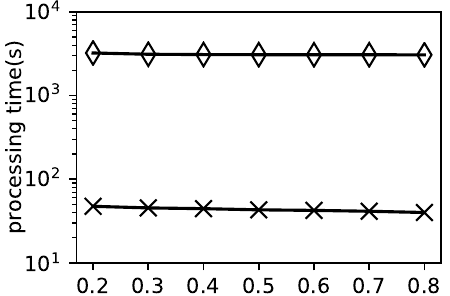}
			\vspace{-0.6cm}
		\end{minipage}
	}
	\vspace{-0.2cm}
	\\
	\caption{Performance when varying $\epsilon$ ($ \mu=6 $)}
	\label{fig:exp1}
\end{figure}

\begin{figure}[htbp]
	\vspace{-0.2cm}
	\centering
	\begin{minipage}[b]{0.25\textwidth}
		\includegraphics[width=1\textwidth]{figure/legend.pdf}
	\end{minipage}
	\vspace{-0.2cm}
	\\
	\subfigure[\kw{EW}]{
		\begin{minipage}[b]{0.20\textwidth}
			\includegraphics[width=1\textwidth]{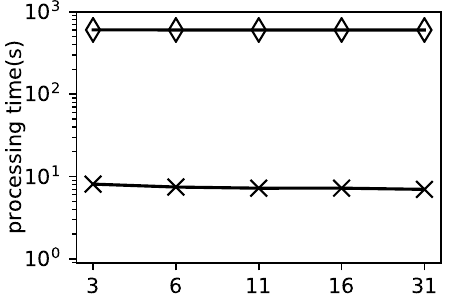}
			\vspace{-0.6cm}
		\end{minipage}
	}
	\subfigure[\kw{CN}]{
		\begin{minipage}[b]{0.20\textwidth}
			\includegraphics[width=1\textwidth]{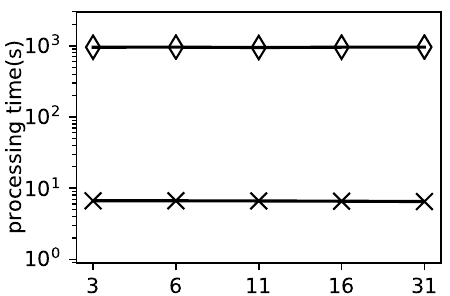}
			\vspace{-0.6cm}
		\end{minipage}
	}
	\vspace{-0.2cm}
	\\
	\subfigure[\kw{HW}]{
		\begin{minipage}[b]{0.20\textwidth}
			\includegraphics[width=1\textwidth]{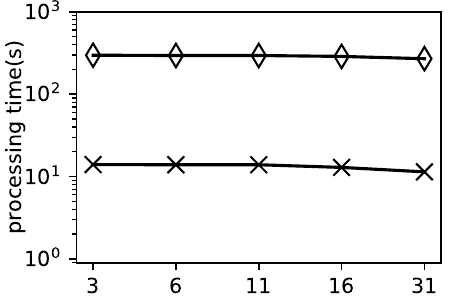}
			\vspace{-0.6cm}
		\end{minipage}
	}
	\subfigure[\kw{OR}]{
		\begin{minipage}[b]{0.20\textwidth}
			\includegraphics[width=1\textwidth]{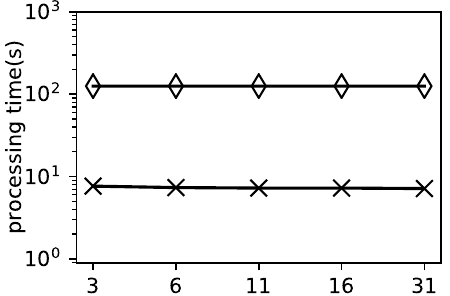}
			\vspace{-0.6cm}
		\end{minipage}
	}
	\vspace{-0.2cm}
	\\    
	\subfigure[\kw{TP}]{
		\begin{minipage}[b]{0.20\textwidth}
			\includegraphics[width=1\textwidth]{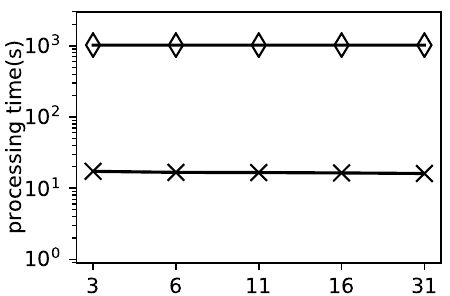}
			\vspace{-0.6cm}
		\end{minipage}
	}
	\subfigure[\kw{UK}]{
		\begin{minipage}[b]{0.20\textwidth}
			\includegraphics[width=1\textwidth]{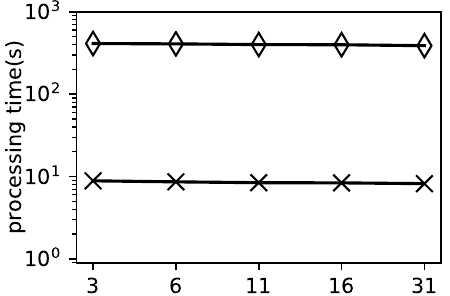}
			\vspace{-0.6cm}
		\end{minipage}
	}
	\vspace{-0.2cm}
	\\    
	\subfigure[\kw{EU}]{
		\begin{minipage}[b]{0.20\textwidth}
			\includegraphics[width=1\textwidth]{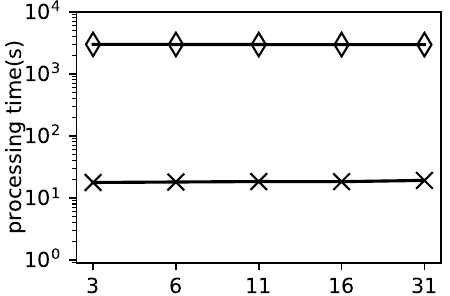}
			\vspace{-0.6cm}
		\end{minipage}
	}
	\subfigure[\kw{ST}]{
		\begin{minipage}[b]{0.20\textwidth}
			\includegraphics[width=1\textwidth]{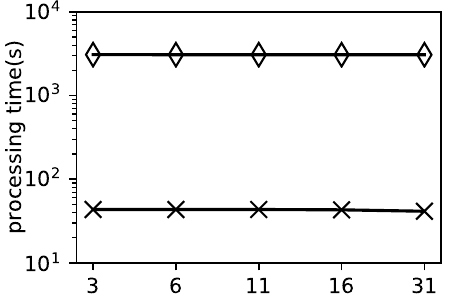}
			\vspace{-0.6cm}
		\end{minipage}
	}
	\vspace{-0.2cm}
	\\
	\caption{Performance when varying $\mu$ ($\epsilon=0.5 $)}
	\label{fig:exp2}
	\vspace{-0.1cm}
\end{figure}

\stitle{Datasets.} We evaluate the algorithms on ten real graphs. \kw{OR} and \kw{FR} are downloaded from the SNAP \cite{snapnets}. \kw{TP} and \kw{ST} are downloaded from the Network Repository \cite{graphrepository2013}. \kw{CN}, \kw{EW}, \kw{HW}, \kw{UK}, and \kw{EU} are downloaded from WebGraph \cite{BoVWFI,BRSLLP,BCSU3}. The details are shown in \reftable{datasets}. The first eight datasets are used to evaluate the in-memory algorithms, while the last two datasets cannot fit into the Nvidia Tesla V100 GPU and are used to evaluate the out-of-core algorithms.

\stitle{Algorithms.} We evaluate the following algorithms:



\begin{itemize}[leftmargin=*]
\item \gpuscan: The state-of-the-art GPU-based algorithm \cite{DBLP:journals/tpds/StovallKA15}, which is introduced in \refsec{gpuscan}. 
\item \gpuscanp: Our proposed GPU-based in-memory algorithm (\refalg{gpuscanp} in \refsec{inmemory}).
\item \gpuscanpuvm: Direct out-of-core  algorithm, namely, \gpuscanp based on the \kw{UVM}.  
\item \gpuscanpo: Our proposed GPU-based out-of-core  algorithm (\refalg{outofcore} in \refsec{outofcore}).	
\end{itemize}

We use {CUDA} 10.1  and {GCC} 4.8.5 to compile all codes with -O3 option. The time cost of the algorithms is measured as the amount of elapsed wall-clock time during the execution. We set the maximum running time for each test to be 100,000 seconds. If a test does not stop within the time limit,  we denote the corresponding running time as \kw{INF}. For $\epsilon$, we choose $\epsilon \in \{0.2, 0.3, 0.4, 0.5, 0.6,0.7,0.8\}$ with $\mu = 6$ as default. For $\mu$, we choose $\mu \in \{3, 6, 11, 16, 31\}$ with $\epsilon = 0.5$ as default.

\begin{table*}[!t]
	\caption{Time consumption of  \gpuscan and \gpuscanp in each phase with $ \epsilon=0.5$ and $\mu=6$ (s)}
	\label{tab:exp}
	\vspace{-0.2cm}
	\begin{center}
		\small{
			\setlength{\tabcolsep}{1.3mm}{
				\begin{tabular}{c|ccc|ccc|ccc}
				\hline
				\multicolumn{1}{c|}{\multirow{2}{*}{Dataset}}   & \multicolumn{3}{c|}{Phase 1 }  &\multicolumn{3}{c|}{Phase 2}  & \multicolumn{3}{c}{Phase 3} \\ \cline{2-10}
				\multicolumn{1}{c|}{} & \gpuscan   & \gpuscanp  & Speedup  & \gpuscan     & \gpuscanp    & Speedup    & \gpuscan & \gpuscanp & Speedup \\ \hline\hline
				\kw{EW}  & 581.32   & 6.35  & 91.55  & 0.80    & 0.022 & 3.64   & 10.60  & 0.23 & 46.08   \\
				\kw{CN}  & 846.88	& 4.31  & 196.49 & 94.50  & 0.025 & 3780.00   & 6.57  & 0.071 & 92.48   \\
				\kw{HW}  & 203.84   & 9.33  & 21.85  & 76.579 & 0.010 & 7657.90 & 5.94  & 0.017 & 349.41    \\
				\kw{OR}  & 108.43   & 6.02  & 18.01  & 1.14  & 0.013 & 87.69   & 8.49  & 0.12 & 70.75  \\
				\kw{TP}  & 998.70   & 13.65 & 73.16  & 4.24   & 0.021 & 201.90   & 10.87  & 1.25 & 8.696   \\
				\kw{UK}  & 315.31   & 6.95  & 45.37  & 65.33  & 0.058 & 1126.38  & 11.96 & 0.16 & 74.75   \\
				\kw{EU}  & 2456.73  & 16.58 & 148.17 & 493.97  & 0.036 & 13721.39  & 13.09  & 0.237 & 55.23 \\ 
				\kw{ST}  & 3030.11	& 40.04 & 75.68  & 11.39  & 0.064 & 117.97   & 18.20 & 0.30 & 60.67   \\
				\hline
			\end{tabular}
			}
		}
	\end{center}
	\vspace{-0.2cm}
\end{table*}

\subsection{Performance Studies on In-memory Algorithms}

\stitle{Exp-1: Performance when varying $\epsilon$.}  In this experiment, we evaluate the performance of \gpuscan and \gpuscanp by varying $\epsilon$. We report the processing time on the first eight datasets in \reffig{exp1}.

\reffig{exp1} shows that \gpuscan consumes much more time when varying the value of  $ \epsilon $. For example, on \kw{EU}, \gpuscan takes 3141.11s to finish the clustering when $ \epsilon=0.2$ while \gpuscanp only takes 18.93s in the same situation, which achieves 168 times speedup. This is because lots of extra costs are introduced in \gpuscan compared with \gpuscanp, which is consistent with the time complexity analysis in \refsec{existing} and \refsec{inmemory}. Moreover, \reffig{exp1} shows that both the running time of \gpuscan and \gpuscanp decrease when the value of $\epsilon$ increases. For \gpuscan, this is because as $\epsilon$ increases, the number of core vertices and clusters decreases, which means the time for cluster detection in \gpuscan decreases. Additionally,  Exp-3 shows that the cluster detection phase accounts for a great proportion of the total time. Therefore, the running time of \gpuscan decreases as $\epsilon$ increases. For \gpuscanp, as the value of $\epsilon$ increases, more vertex pairs are dissimilar. Consequently, more unnecessary similarity computation is avoided. Thus, the running time decreases as well.


\stitle{Exp-2: Performance when varying $\mu$.} In this experiment, we evaluate the performance of \gpuscan and \gpuscanp by varying $ \mu $. The results are shown in \reffig{exp2}.

As shown in \reffig{exp2}, \gpuscanp is much more efficient than \gpuscan. For example, int the \kw{UK}, \gpuscan takes 411.97s to finish the clustering when $ \mu=3 $ while \gpuscanp only takes 8.866s. The reasons are the same as discussed in Exp-1. 


\stitle{Exp-3: Time consumption of each phase.} In this experiment, we compare the time consumption of \gpuscan and \gpuscanp in each phase with $ \epsilon=0.5$ and $\mu=6$. And the results are shown in \reftable{exp}.

As shown in \reftable{exp}, \gpuscanp achieves significant speedup in all  three phases compared with \gpuscan, especially for Phase 2. For example, on dataset \kw{CN}, \gpuscan takes 846.88s/94.50s/6.57s in Phase 1/Phase 2/Phase 3, respectively. On the other hand, the consuming time of \gpuscanp for these three phases is 4.31s/0.025s/0.071s, which achieves 196.49/3780.00/92.48 times speedup, respectively. The reasons are as follows: for  Phase 1, the span of \gpuscan is $O(\mydeg_{\kw{max}})$, while \gpuscanp reduces the span to $O(\log(\mydeg_{\kw{max}}))$. For Phase 2, \gpuscan has to construct $E'.u$ and $E'.v$ by $\kw{sort()}$ and $\kw{partition}()$ in \kw{Thrust} library in each iteration, which is time-consuming. Moreover, it needs several iterations to finish the whole cluster detection, which is shown in the last column of \reftable{datasets}. On the other hand, for \gpuscanp, we construct the spanning forest based on the \kw{parent} array directly and the time-consuming retrieval of $E'.u$ and $E'.v$ is avoided due to the CSR-Enhanced graph layout structure. For phase 3, \gpuscan still needs to retrieve  $E'.u$ and $E'.v$ from  $A.u$ and $A.v$ while \gpuscanp totally avoid these time-consuming operation. 


%

\subsection{Performance Studies on Out-of-Core Algorithms}

\begin{figure}[htbp]
	\vspace{-0.2cm}
	\centering
	\begin{minipage}[b]{0.28\textwidth}
		\includegraphics[width=1\textwidth]{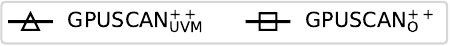}
	\end{minipage}
	\vspace{-0.2cm}
	\\
	\subfigure[\kw{TW}]{
		
		\includegraphics[width=1.42in]{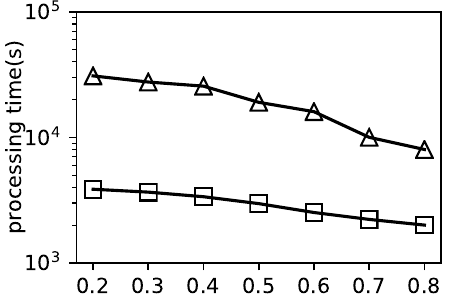}
		\label{twitter-2010}
	}
	\subfigure[\kw{FR}]{
		\includegraphics[width=1.42in]{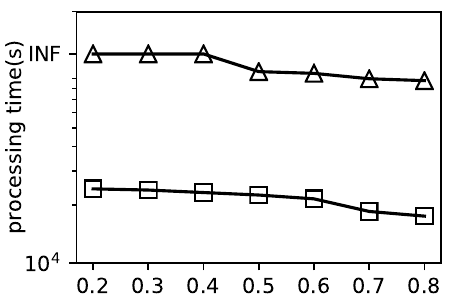}
		\label{comfriendster}
	}
	\vspace{-0.2cm}
	\caption{Performance when varying $\epsilon$ ($\mu = 6$)}
	\label{fig:exp4}
	\vspace{-0.2cm}
\end{figure}

\stitle{Exp-5: Performance when varying $\epsilon$.} In this experiment, we evaluate the performance of \gpuscanpuvm and \gpuscanpo when varying the value of $\epsilon$ on datasets \kw{TW} and \kw{FR}. The results are shown in \reffig{exp4}.

\reffig{exp4} shows that \gpuscanpo is much more efficient than \gpuscanpuvm on these two datasets. For example, on \kw{TW}, \gpuscanpuvm takes 30886.315s to finish the clustering when $\epsilon = 0.2$ while \gpuscanpo finishes the clustering in 3866.424s. On \kw{FR}, \gpuscanpuvm cannot finish the clustering when $\epsilon = 0.2, 0.3, 0.4$ while \gpuscanpo finishes the clustering in a reasonable time. This is because UVM aims to provide a general mechanism to overcome the GPU memory limitation. However, as the data locality of structure clustering is poor, \gpuscanpuvm involves lots of data movements between CPU memory and GPU memory, which is time-consuming. On the other hand, \gpuscanpo conducts the clustering based on the edge extended subgraph, which not only avoids the time-consuming data movements but also overcomes the GPU memory limitation. Moreover, the running time of \gpuscanpo and \gpuscanpuvm decreases as the value of $\epsilon$ increases. This is because as the value of $\epsilon$, more vertex pairs are dissimilar. Consequently, more unnecessary similarity computation is avoided in these two algorithms. 

\begin{figure}[h]
	\vspace{-0.2cm}
	\centering
	\begin{minipage}[b]{0.28\textwidth}
		\includegraphics[width=1\textwidth]{figure/legend3.pdf}
	\end{minipage}
	\vspace{-0.2cm}
	\\
	\subfigure[\kw{TW}]{
		
		\includegraphics[width=1.42in]{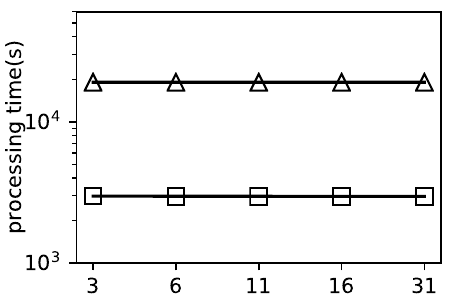}
		\label{twitter-2010-2}
	}
	\subfigure[\kw{FR}]{
		\includegraphics[width=1.42in]{figure/exp5-tw.pdf}
		\label{comfriendster-2}
	}
	\vspace{-0.2cm}
	\caption{Performance when varying $\mu$ ($\epsilon = 0.5$)}
	\label{fig:exp5}
	\vspace{-0.2cm}
\end{figure}

\stitle{Exp-6: Performance when varying $\mu$.} In this experiment, we evaluate the performance of \gpuscanpuvm and \gpuscanpo when varying the value of $\mu$ on dataset \kw{TW} and \kw{FR}. The results are shown in \reffig{exp5}.

As shown in \reffig{exp5}, \gpuscanpo is much more efficient than \gpuscanpuvm. For example, on \kw{TW}, \gpuscanpuvm takes 19037.683s to finish the clustering when $\mu = 3$ while \gpuscanpo finishes the clustering in 2974.947s. The reasons are the same as discussed in Exp-5.

\section{Conclusion}
In this paper, we study the GPU-based structural clustering problem. Motivated by the state-of-the-art GPU-based structural clustering algorithms that suffer from inefficiency and GPU memory limitation,  we propose new GPU-based structural clustering algorithms. For efficiency issues, we propose a new progressive clustering method tailored for GPUs that not only avoids extra parallelization costs but also fully exploits the computing resources of GPUs. To address the GPU memory limitation issue, we propose a partition-based algorithm for structural clustering that can process large graphs with limited GPU memory. We conduct experiments on ten real graphs and the experimental results demonstrate the efficiency of our proposed algorithm.


\bibliographystyle{abbrv}
\bibliography{ref.bib}

\vspace{-2cm}
\begin{IEEEbiography}[{\includegraphics[width=1in,height=1.25in,clip,keepaspectratio]{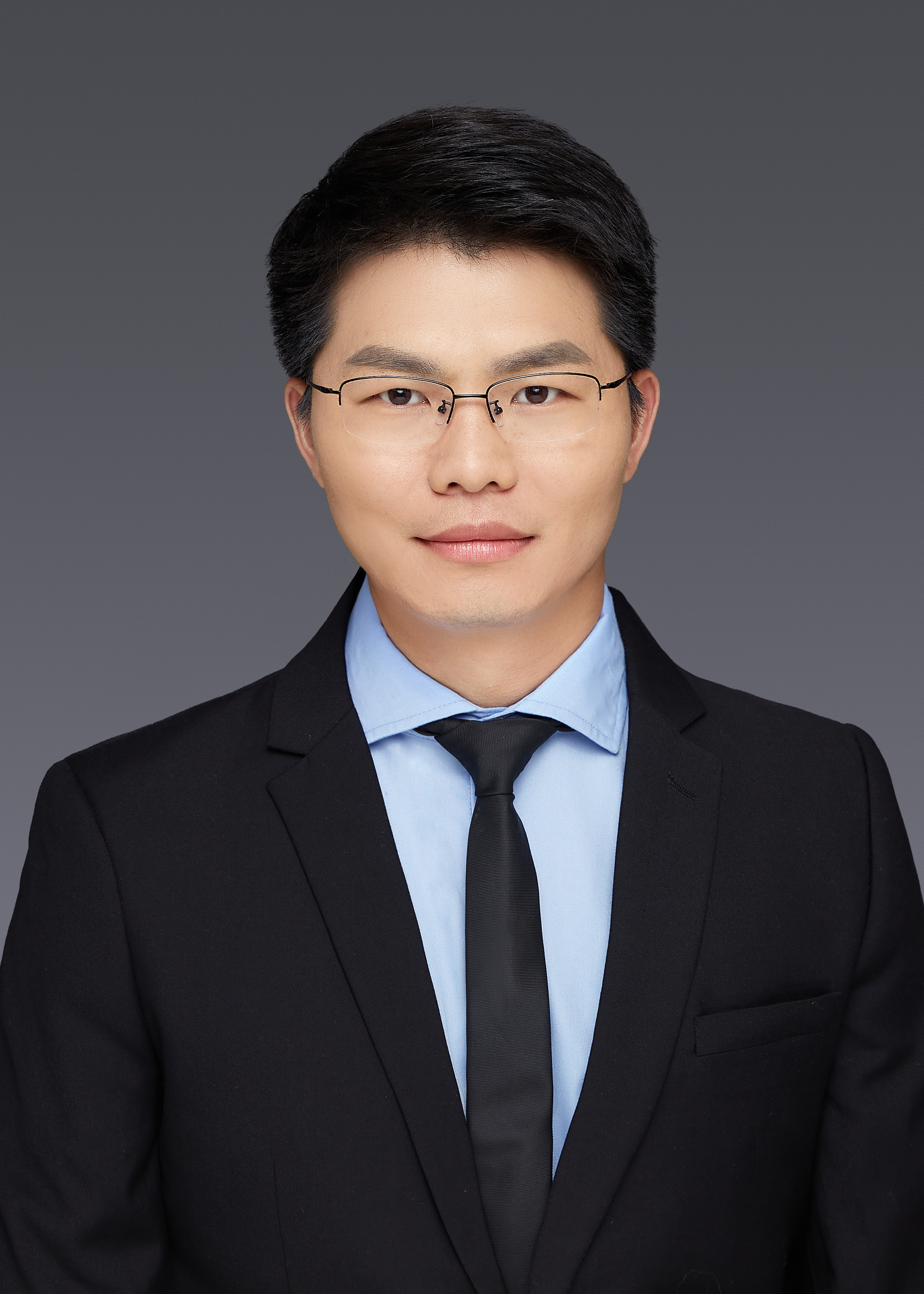}}]{Long Yuan} is currently a professor in School of Computer Science and Engineering, Nanjing University of Science and Technology, China. He received his PhD degree from the University of New South Wales, Australia, M.S. degree and B.S. degree both from Sichuan University, China. His research interests include graph data management and analysis.  He has published papers in conferences and journals including VLDB, ICDE, WWW, The VLDB Journal, and TKDE.
\end{IEEEbiography}
\vspace{-2.0cm}

\begin{IEEEbiography}[{\includegraphics[width=1in,height=1.25in,clip,keepaspectratio]{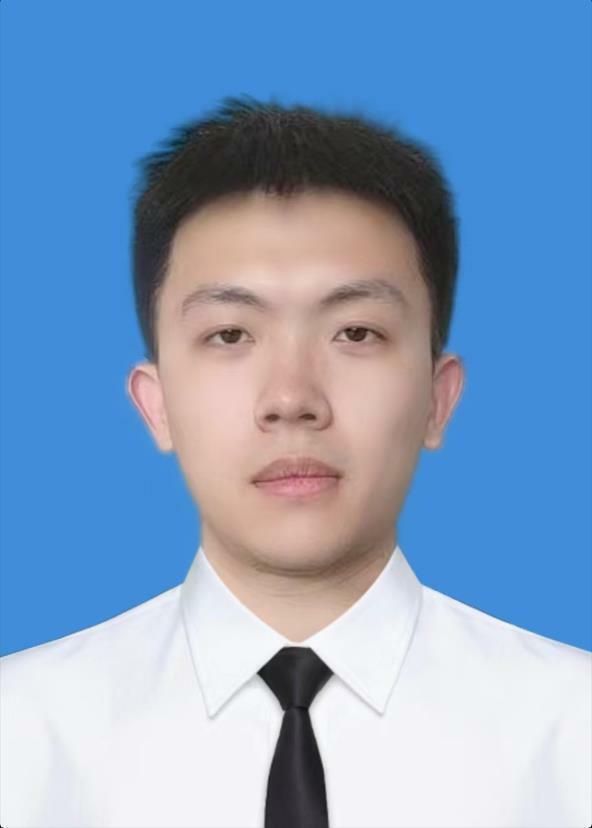}}]{Zeyu Zhou} Zeyu Zhou is currently a master student in the Department of Computer Science, Nanjing University of Science and Technology. His research interests include graph data management and analysis.
\end{IEEEbiography}
\vspace{-7cm}

\begin{IEEEbiography}[{\includegraphics[width=1in,height=1.25in,clip,keepaspectratio]{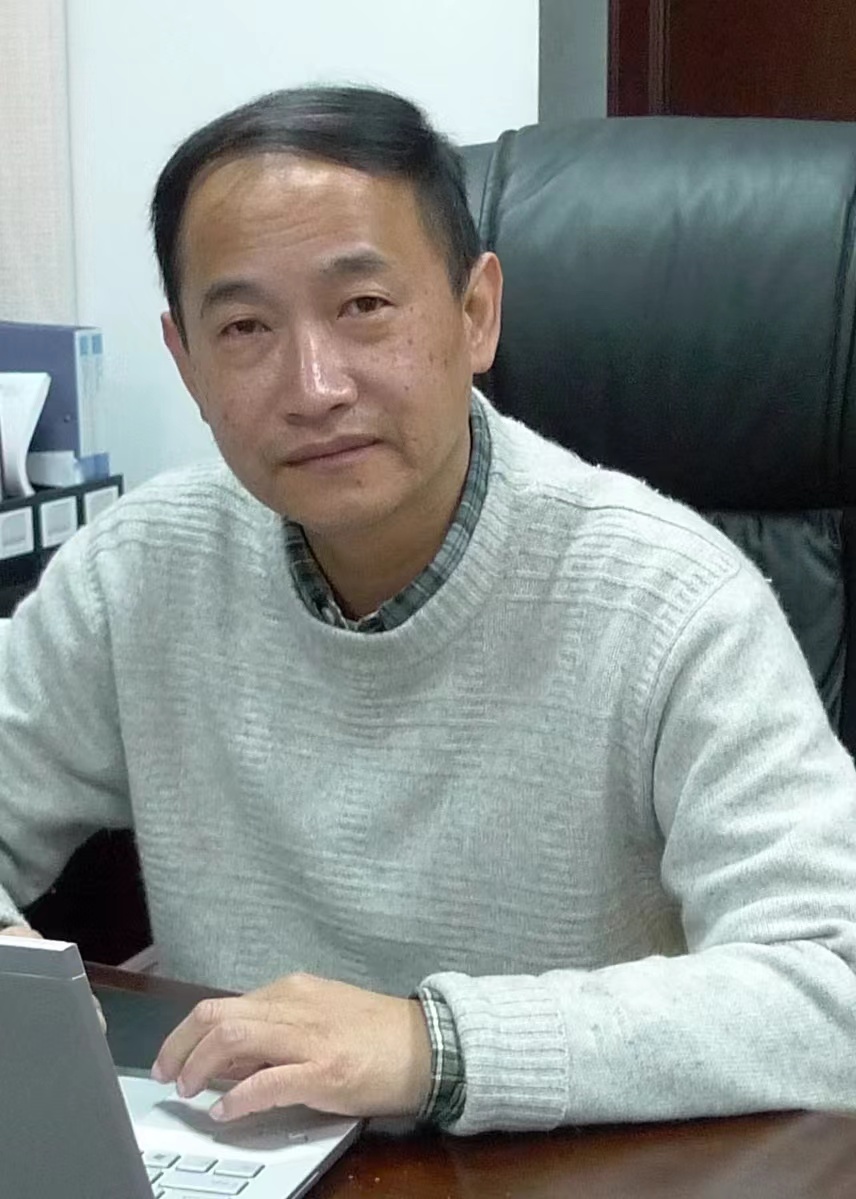}}]{Xuemin Lin} received the BSc degree in applied math from Fudan University, in 1984, and the PhD degree in computer science from the University of Queensland, in 1992. He is a professor with the School of Computer Science and Engineering, University of New South Wales. His current research interests include approximate query processing, spatial data analysis, and graph visualization. He is a fellow of the IEEE.
\end{IEEEbiography} 
\vspace{-7.5cm}

\begin{IEEEbiography}[{\includegraphics[width=1in,height=1.25in,clip,keepaspectratio]{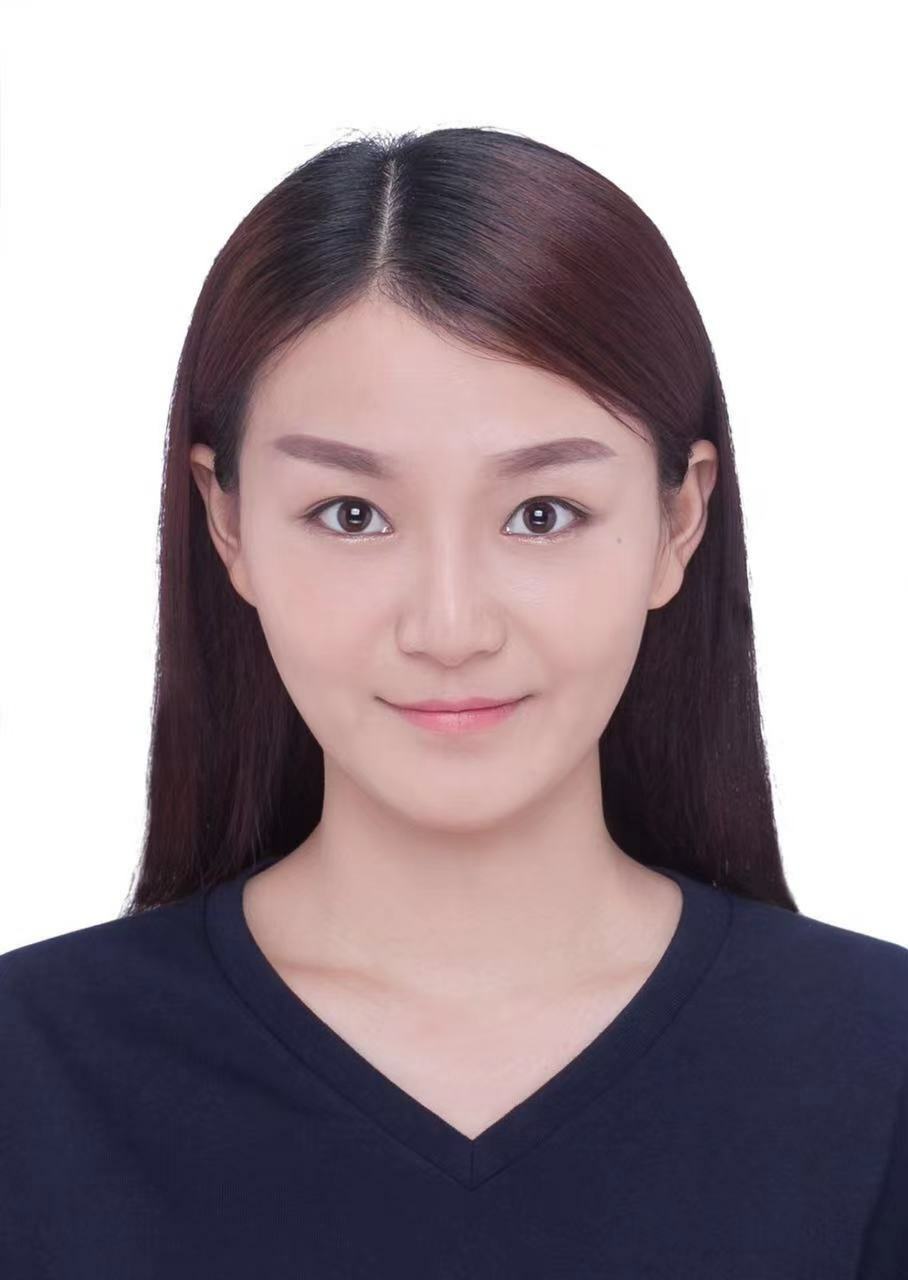}}]{Zi Chen} received the PhD degree from the Software Engineering Institute, East China Normal University, in 2021. She is currently a postdoctor in ECNU. Her research interest is big graph data query and analysis, including cohesive subgraph search on large-scale social networks and path planning on road networks.
\end{IEEEbiography}

\vspace{-7.5cm}

\begin{IEEEbiography}
[{\includegraphics[width=1in,height=1.25in,clip,keepaspectratio]{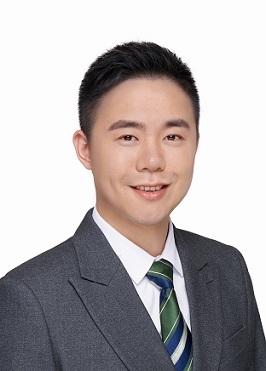}}]{Xiang Zhao} received the PhD degree from The University of New South Wales, Australia, in 2014. He is currently a professor with the National University of Defense Technology, China. His research interests include graph datamanagement and mining. He is a member of  the IEEE. He has published papers in conferences and journals including VLDB, ICDE, WWW, The VLDB Journal, and TKDE.
\end{IEEEbiography}

\vspace{-7.5cm}

\begin{IEEEbiography}[{\includegraphics[width=1in,height=1.25in,clip,keepaspectratio]{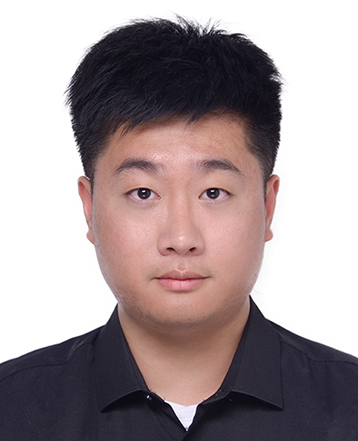}}]
{Fan Zhang} is a Professor at Guangzhou University. He was a research associate in the School of Computer Science and Engineering, University of New South Wales, from 2017 to 2019. He received the BEng degree from Zhejiang University in 2014, and the PhD from University of Technology Sydney in 2017. His research interests include graph algorithms and social networks. Since 2017, he has published more than 20 papers in top venues, e.g., SIGMOD, PVLDB, ICDE, IJCAI, AAAI, TKDE and VLDB Journal.
\end{IEEEbiography}

\end{document}